\documentclass[11pt,numberedappendix]{emulateapj}
\usepackage{epsfig} 
\slugcomment{To appear in ApJS} 
\shorttitle{Faint Galaxies in Deep ACS Observations} 
\shortauthors{Ben\'\i tez et al.}

\begin{document}

\title{Faint Galaxies in Deep ACS observations}

\author{
N. Ben\'\i tez\altaffilmark{1}, 
H. Ford\altaffilmark{1}, 
R. Bouwens\altaffilmark{2}, 
F. Menanteau \altaffilmark{1}, 
J. Blakeslee\altaffilmark{1}, 
C. Gronwall\altaffilmark{3}, 
G. Illingworth \altaffilmark{2}, 
G. Meurer \altaffilmark{1},
T.J.Broadhurst\altaffilmark{4},
M. Clampin\altaffilmark{5},
M. Franx \altaffilmark{6},
G.F.Hartig\altaffilmark{5},
D. Magee\altaffilmark{2},
M. Sirianni\altaffilmark{1}, 
D.R. Ardila\altaffilmark{1}
F. Bartko\altaffilmark{7}, 
R.A. Brown\altaffilmark{5},
C.J. Burrows\altaffilmark{5},
E.S. Cheng\altaffilmark{8},
N.J.G.Cross\altaffilmark{1},
P.D. Feldman\altaffilmark{1},
D.A. Golimowski\altaffilmark{1},
L.   Infante\altaffilmark{9}
R.A. Kimble\altaffilmark{8},
J.E. Krist\altaffilmark{5},
M.P. Lesser\altaffilmark{10},
Z.   Levay \altaffilmark{5},
A.R. Martel\altaffilmark{1},
G.K. Miley\altaffilmark{6},
M.   Postman\altaffilmark{5},
P.   Rosati\altaffilmark{11}, 
W.B. Sparks\altaffilmark{5}, 
H.D. Tran\altaffilmark{1}, 
Z.I. Tsvetanov\altaffilmark{1},   
R.L. White\altaffilmark{1,5}
\& W. Zheng\altaffilmark{1}
}

\altaffiltext{1}{Department of Physics and Astronomy, Johns Hopkins
University, 3400 North Charles Street, Baltimore, MD 21218.}

\altaffiltext{2}{UCO/Lick Observatory, University of California, Santa
Cruz, CA 95064.}

\altaffiltext{3}{Department of Astronomy and Astrophysics, The
Pennsylvania State University, 525 Davey Lab, University Park, PA
16802.}

\altaffiltext{4}{Racah Institute of Physics, The Hebrew University,
Jerusalem, Israel 91904.}

\altaffiltext{5}{STScI, 3700 San Martin Drive, Baltimore, MD 21218.}

\altaffiltext{6}{Leiden Observatory, Postbus 9513, 2300 RA Leiden,
Netherlands.}

\altaffiltext{7}{Bartko Science \& Technology, P.O. Box 670, Mead, CO
80542-0670.}

\altaffiltext{8}{NASA Goddard Space Flight Center, Laboratory for
Astronomy and Solar Physics, Greenbelt, MD 20771.}

\altaffiltext{9}{Departmento de Astronom\'{\i}a y Astrof\'{\i}sica,
Pontificia Universidad Cat\'{\o}lica de Chile, Casilla 306, Santiago
22, Chile.}

\altaffiltext{10}{Steward Observatory, University of Arizona, Tucson,
AZ 85721.}

\altaffiltext{11}{European Southern Observatory,
Karl-Schwarzschild-Strasse 2, D-85748 Garching, Germany.}

\begin{abstract}
  We present the analysis of the faint galaxy population in the
Advanced Camera for Surveys (ACS) Early Release Observation fields VV 29 
(UGC 10214) and NGC 4676. These observations cover a total area of
26.3 arcmin$^2$, and have depths close to that of the Hubble Deep
Fields in the deepest part of the VV 29 image, with $10\sigma$ 
detection limits for point sources of $27.8$, $27.6$ and $27.2$ 
AB magnitudes in the $g_{F475W}$, $V_{F606W}$ and $I_{F814W}$ 
bands respectively. 

Measuring the faint galaxy number count distribution is 
a difficult task, with different groups arriving at
widely varying results even on the same dataset. Here we attempt to
thoroughly consider all aspects relevant for faint galaxy counting and
photometry, developing methods which are based on public software 
and that are easily reproducible by other astronomers. Using simulations we
determine the best $SExtractor$ parameters for the detection of faint
galaxies in deep HST observations, paying special attention to the
issue of deblending, which significantly affects the normalization and
shape of the number count distribution.  We confirm, as claimed 
by Bernstein, Freedman and Madore (2002a,b; hereafter BFM), that Kron-like 
magnitudes, such as the ones generated by $SExtractor$, can miss more
than half of the light of faint galaxies, what dramatically affects the slope 
of the number counts. 
We show how to correct for
this effect, which depends sensitively not only on the characteristics 
of the observations, but also on the choice of $SExtractor$ parameters. 

  We present catalogs for the VV 29 and NGC 4676 fields with photometry 
in the $F475W, F606W$ and $F814$ bands. We also show that combining the 
Bayesian software BPZ with superb ACS data and new spectral templates 
enables us to estimate reliable photometric redshifts for a significant 
fraction of galaxies with as few as three filters. 

 After correcting for selection effects, we measure slopes of 
$0.32\pm 0.01$ for $22< g_{F475W}< 28$, 
$0.34\pm 0.01$ for $22<V_{F606W}<27.5$ and $0.33\pm 0.01$ for 
$22<m_{F814W}<27$. The counts do not flatten (except perhaps in the 
$F475W$ filter), up to the depth of our observations. Our results agree 
well with those of BFM, 
who used different datasets and techniques, and show that it is possible 
to perform consistent measurements of galaxy number counts if the selection 
effects are properly considered. We find that the faint counts 
$m_{AB}> 25.5$ can be well approximated in all our filters by a passive 
luminosity evolution model based on the COMBO-17 luminosity function 
($\alpha=-1.5$) , with a strong merging rate following the prescription of 
Glazebrook et al. (1994), $\phi^*\propto (1+Qz)$, with $Q=4$. 

\end{abstract}


\section{Introduction}

  On March 7th 2002, the Advanced Camera for Surveys (ACS, Ford et
al. 1998, 2002) was installed in HST during the space shuttle mission
ST-109. ACS is an instrument designed and built with the study of the
faint galaxy population as one of its main goals. Here we describe the
processing and analysis of some of the first science observations
taken with the ACS Wide Field Camera, called Early Release 
Observations (EROs).
 
  An important result obtained with the WFPC2 observations of the
Hubble Deep Fields (Williams et al. 1996, Casertano et al. 2000) was
the measurement of the galaxy number count distribution to very faint
\( (m_{AB}\gtrsim 27) \) limits. However, it is remarkable that
different groups have reached different conclusions about the slope
and normalization of the number counts even when using the same software on
the same dataset (see e.g. Ferguson, Dickinson \& Williams 2000,
Vanzella et al. 2001).  One of the few results on which all groups
seemed to agree was the flattening of the number counts at $I_{AB}\sim
26$. However, BFM claim that this is a spurious effect, caused by 
the underestimation of the true luminosity of faint galaxies by standard 
aperture measurements, and that the number counts continue with a slope of
$0.33$ up to the limits of the HDFN V and I bands.

 Two of the main goals of this paper are improving our understanding
of the biases and selection effects involved in counting and measuring
the properties of very faint galaxies, and developing techniques that
can be applied to a wide range of observations and that are easily
reproduced by other astronomers. This is essential if galaxy counting
is to become a precise science. We have done this by using almost
exclusively public software, and specifying the parameters used,
thus ensuring that our results are repeatable.
 
  Our final results are the number counts in the $F475W$, $F606W$ and
$F814W$ bands, carefully corrected for selection effects. Using an
independent procedure, we confirm the apparent absence of flattening
in the number counts found by BFM.  
We also show that using proper priors, reasonably robust
photometric redshifts can be obtained using only three ACS
filters. Finally, we present photometric catalogs of field galaxies in
the VV 29 and NGC 4676 fields.

The structure of the paper is the following: Section 2 describes our 
observations,  
Section 3 deals with the image processing and the generation of the catalogs, 
including the description of the simulations used to correct our number 
counts. Section 4 lists and explains the quantities included in our catalogs, 
Section 5 presents our number counts, and Section 6 summarizes our main 
results and conclusions.

\section{Observations}

The observations analyzed here were obtained with the Wide Field
Camera of the Advanced Camera for Surveys (Ford et al. 1998, 2002) and include
two fields.  The first is centered on VV 29 (Vorontsov-Velyaminov
1959), also known as UGC 10214 and Arp 188 (Arp 1966), a
bright spiral with a spectacular tidal tail at $z=0.032$.  Due to a
pointing error, the field was imaged twice, resulting in a central
region with twice the exposure time of the NGC 4676 field. The galaxy
itself and its associated star formation has been considered in detail
by Tran et al.(2002).  The second field is centered on NGC 4676
(Holmberg, 1937), an interacting galaxy pair at $z=0.022$. Figures 1
and 2 show the ACS images of these fields, and Table 1 summarizes the
main characteristics of the observations.

\begin{figure}[ht]
\epsfig{file=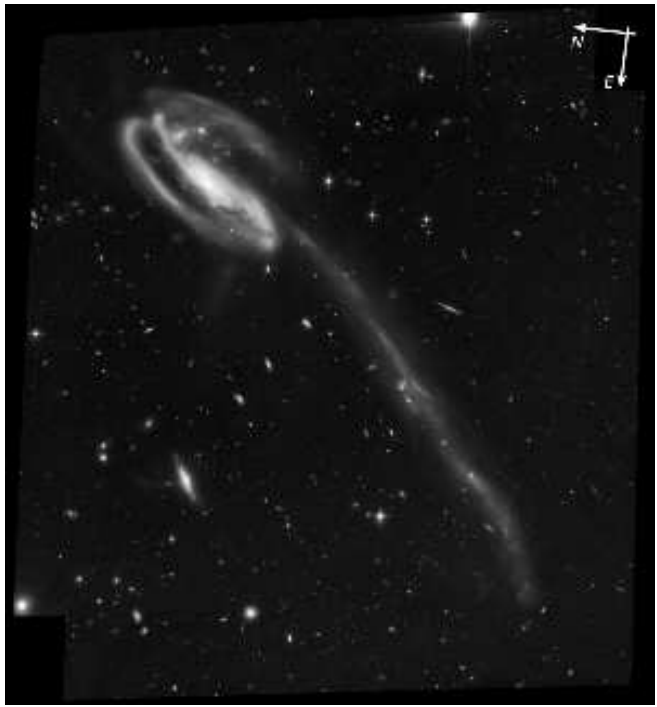,width=\linewidth}
\caption{ Color image of VV 29, the ``Tadpole'', 
obtained by combining the ACS WFC F475W, F606W and F814W filters. 
The observations are described in Table 1.
}
\end{figure}

\begin{figure}[ht]
\epsfig{file=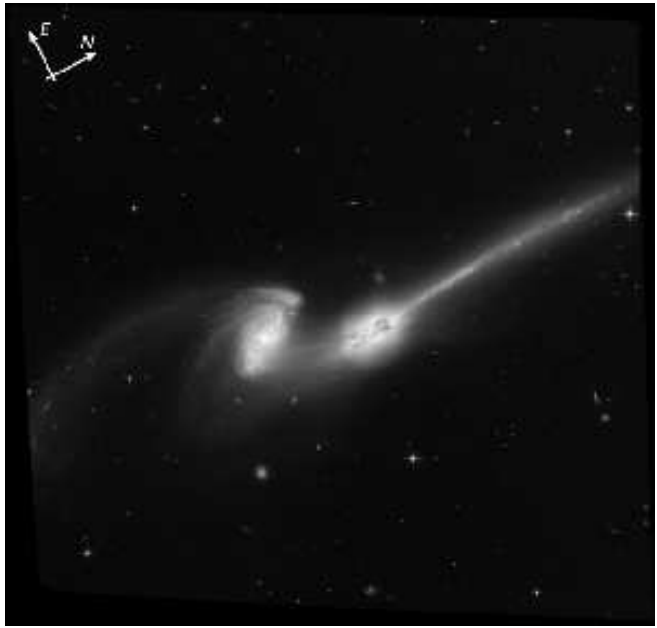,width=\linewidth}
\caption{Color image of NGC 4676, the ``Mice'' 
obtained by combining the ACS WFC F475W, F606W and F814W filters. 
The observations are described in Table 1.
}
\end{figure}

\begin{deluxetable*}{llllccc}
\tabletypesize{\scriptsize} 
\tablecaption{Early Release Observations of VV 29(UGC 10214) and NGC 4676} 
\tablewidth{0pt} 
\tablehead{
\colhead{Field} & 
\colhead{RA(J2000)} & 
\colhead{DEC(J2000)} &
\colhead{ACS WFC filter} & 
\colhead{Exposure time} & 
\colhead{N exposures} & 
\colhead{Area (arcmin$^2$)} 
} 
\startdata 
VV 29 & 16:06:17.4 & 55:26:46 & F475W & $13600$ & $12$ & 14.48 \\ 
VV 29 & 16:06:17.4 & 55:26:46 & F606W & $8040 $ & $12$ & 14.49 \\
VV 29 & 16:06:17.4 & 55:26:46 & F814W & $8180 $ & $12$ & 14.46 \\ 
NGC 4676 & 12:46:09.0 & 30:44:25 & F475W & $6740 $ & $6$ & 11.84 \\
NGC 4676 & 12:46:09.0 & 30:44:25 & F606W & $4000 $ & $6$ & 11.84 \\
NGC 4676 & 12:46:09.0 & 30:44:25 & F814W & $4070 $ & $6$ & 11.84 \\ 
\enddata

\end{deluxetable*}

\section{Data analysis}

\subsection{Image processing}

   A brief description of the calibration and reduction procedures for
the VV 29 field can be found in Tran et al. (2003).  
The raw ACS data were processed through the standard CALACS pipeline 
(Hack 1999) at STScI.  This included overscan, bias, and dark subtraction, 
as well as flat-fielding.  CALACS also converts the image counts to
electrons and populates the header photometric keywords.  About half
of the images in these datasets were taken as cosmic ray (CR)
split pairs that were combined into single ``crj'' images by
CALACS; the rest were taken as single exposures.

The calibrated images were then processed through the ``Apsis'' ACS
Investigation Definition Team pipeline, described in detail by
Blakeslee et al.\ (2003).  Briefly, Apsis finds all bright compact
objects in the input images, sorts through the catalogs to remove the
cosmic rays and obvious defects, corrects the object positions using
the ACS distortion model (Meurer et al.\ 2003), and then derives the
offsets and rotations for each image with respect to a selected
reference image.  For the present data sets, over one hundred objects
were typically used in deriving the transformation for each image, and
the resulting alignment errors were about 0.04 pix in each direction.
The relative rotation between the first and second epoch VV 29
observations was found to be 0\fdg12.  The offsets and rotations were
then used in combining the individual frames to produce single
geometrically corrected images for each bandpass.

Image combination in Apsis is done with the drizzle software written
by R.\ Hook (Fruchter \& Hook 2002).  The data quality arrays enable
masking of known hot pixels and bad columns, while cosmic rays and
other anomalies are rejected through the iterative drizzle/blot
technique described by Gonzaga et al.\ (1998).  For these
observations, we used the ``square'' (linear) drizzle kernel with an
output scale of 0\farcs05\,pix$^{-1}$. The full width at half maximum
(FWHM) of the point spread function (PSF) was about 0\farcs105, or 2.1
WFC pixels. The linear drizzling of course correlates the noise in 
adjacent pixels,
decreasing the root-mean-squared (RMS) noise fluctuations per pixel by
a factor $1-\frac{1}{3l}$ for our parameters, where $l$ is the linear 
size of the area in which the fluctuations are measured 
(Casertano et al. 2000). However, Apsis calculates RMS arrays for 
each drizzled 
image, i.e., the expected RMS noise per pixel in the absence of 
correlation. These arrays are used later on for image detection, photometric 
noise estimation, etc.  

\begin{figure}[ht]
\epsfig{file=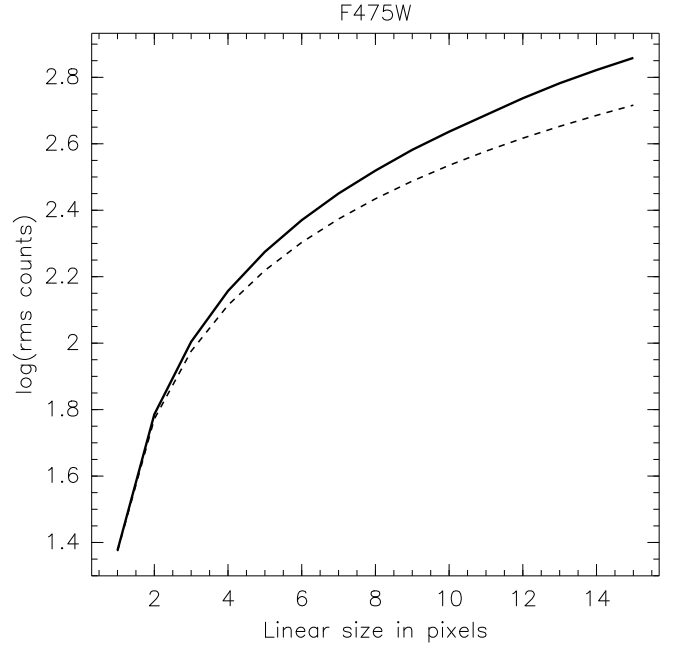,width=\linewidth}
\caption{Behavior of the empirically measured noise in apertures of
varying size (solid line) vs. the expected one (dashed line), for the
central part of the VV 29 F475W images, based on the drizzle parameters
and the noise model of Casertano et al. 2000. As in the case of the
HDFN, the measured noise is higher than the prediction. Similar
behavior is observed in the other filters.
\label{noise}}
\end{figure}

Figure \ref{noise} shows the behavior of the noise
as a function of the size of the area in which it is measured. We see that
it follows well the predicted behavior, but it is slightly higher 
on large scales, an effect which was also noted in the HDFS by Casertano et
al. 2000, and is probably due to intrinsically correlated fluctuations 
in the background galaxy density. 

In addition Apsis detects and performs photometry of stars and
galaxies in the images using $SExtractor$ (Bertin \& Arnouts 1996) and
obtains photometric redshifts for galaxies using the software BPZ
(Ben\'\i tez 2000), steps which will be described in detail below.

The stellar FWHM of our images, \( \sim 0\farcs105 \) is significantly
better than that of WFPC2 observations (e.g., \( \sim 0\farcs14 \) for
the HDFS). A detailed analysis of the ACS WFC point spread function
(PSF) will be published elsewhere (Sirianni et al., in preparation).
Table 2 shows the $10\sigma$ limiting magnitudes for the deep central
area of VV 29 (which was imaged twice), and for the outer area of the
VV 29, together with the NGC 4676 field.

 The absolute accuracy of the positions derived from the information in 
the ACS image header is limited to $\sim 1\arcsec$ by the uncertainty in 
the guide star positions and the alignment of the ACS WFC to Hubble's 
Fine Guidance Sensors.  As a last step, we correct the astrometry of 
the images using the software wcstools and the Guide Star Catalog II 
(Mink et al. 2002).  Although there are not many cataloged stars in 
our images, visual inspection shows that our corrected positions should 
be accurate to $\lesssim 0\farcs1$.

\begin{deluxetable*}{lccc}
\tabletypesize{\scriptsize} \tablecaption{Depth of the VV 29 and NGC
4676 fields compared with the HDFS} \tablewidth{0pt} \tablehead{
\colhead{ } & 
\colhead{VV 29} & 
\colhead{VV 29/NGC 4676} &
\colhead{HDFS} } 
\startdata 

F475W & 27.61 (27.83) & 27.23 (27.45) & 27.97 (27.90) \cr 
F606W & 27.42 (27.64) & 27.09 (27.31) & 28.47 (28.40) \cr 
F814W & 26.98 (27.20) & 26.58 (26.80) & 27.84 (27.77) \cr

\enddata
\tablecomments{Limiting magnitudes for our fields and the HDFS. In each 
entry, the numbers on the left represent the expected $10\sigma$ 
fluctuation in an 0.2 arcsec$^2$ square aperture. The number in parenthesis 
corresponds to the same quantity, but within a circular aperture that has a
diameter 4 times the FWHM of the PSF. We use a value of $0\farcs105$ arcsec for the WFC 
observations and $0\farcs135$ for the HDFS (Casertano et al. 2000). The circular
apertures allow a realistic comparison of the limiting magnitudes for point sources.  
The ACS stellar limiting magnitude through the $4\times$ FWHM aperture ($d \sim 0\farcs42$) is 
$\sim 0.22$ magnitudes {\it fainter} than a source filling the $0\farcs45 \times 0\farcs45$ 
rectangular aperture, whereas the equivalent WFC2 stellar limiting magnitude is 
$\sim 0.07$ magnitudes {\it brighter} because of the WFC2's broader PSF. }

\end{deluxetable*}

 The reduced images in the three filters, together with auxiliary
images (detection image, rms images) can be downloaded from 
http://acs.pha.jhu.edu. 

\subsection{Galaxy identification and photometry}

\subsubsection{Object detection}

$SExtractor$ (Bertin \& Arnouts 1996) has become the \emph{de facto}
standard for automated faint galaxy detection and photometry.  It
finds objects using a connected pixel approach, including weight and
flag maps if desired, and provides the user with efficient and
accurate measurements of the most widely used object
properties. As stated previously, one of our main goals is
to understand in detail how the process of galaxy detection
and analysis affect the shape of the number counts distribution, and then 
use this understanding to arrive at results that are as objective as 
possible.  
To achieve our goal, we carefully choose our $SExtractor$ parameters, and most 
importantly, characterize the biases and errors by using extensive simulations 
with the public software $BUCS$ (Bouwens Universe Construction Set, 
see Appendix A, 
and Bouwens, Broadhurst, and Illingworth 2003 for a detailed description). 
This approach will allow other
astronomers to contrast and compare our results with their own 
in a consistent way. Because the output of $SExtractor$ sensitively
depends on its input parameters, we present our parameters in
Appendix B to ensure that others can repeat our analysis.

One of $SExtractor$'s more convenient features is the double image
mode. This mode enables object detection and aperture definition in
one image, and aperture photometry in a different image. To create our
detection image, we use an inverse variance weighted average of the
$F475W$, $F606W$ and $F814W$ images. This differs from the procedure
followed for the HDFs by other authors, who usually only use the
reddest bands, $F606W$ and $F814W$. However, we think that inclusion
of the $F475W$ image is justified since it is the deepest of the
three; in fact, Table 2 shows that the $F475W$ image is almost as deep
as the HDFS for point sources. The PSF in the final detection image is
basically identical to that of the $F606W$ image, and differs by less
than $2\%$ from the stellar width of the $B$ and $I$ filters.

\begin{figure}[ht]
\epsfig{file=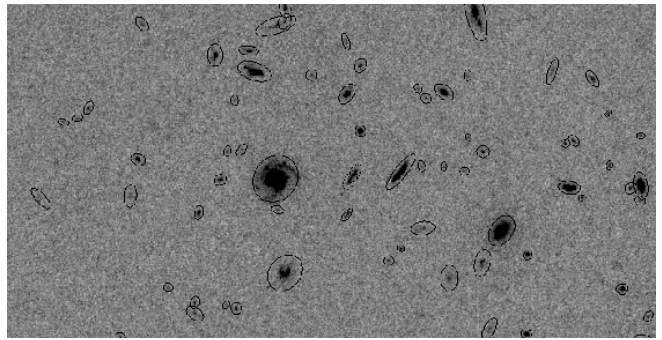,width=\linewidth}
\caption{ Example of $SExtractor$ object finding and deblending on a
section of the VV 29 field. The displayed apertures are the ones corresponding 
to $SExtractor$ $MAG\_AUTO$ magnitudes.\label{SE}}
\end{figure}

\begin{figure}[ht]
\epsfig{file=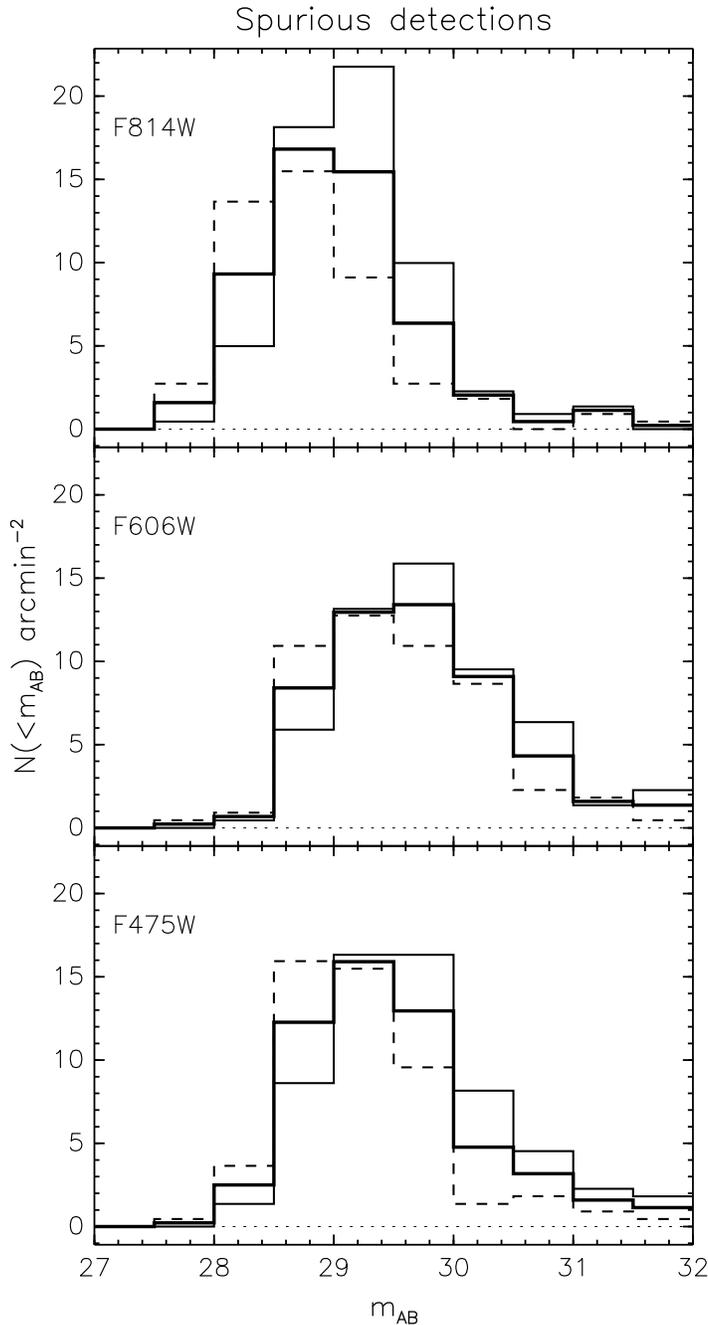}
\caption{Number of expected spurious detections, caused by noise  caused by noise 
fluctuations, as a function of magnitude (see text). 
The thin solid line corresponds to the central part of the VV 29 field 
(see Table 1.), the dashed line to the outer VV 29 field and the NGC 4676 field 
and the thick solid line to whole field.
\label{spurious}}

\end{figure}

Although there are roughly thirteen parameters which influence the
detection process in $SExtractor$, the most critical ones are
$DETECT\_MINAREA$, the minimum number of connected pixels and
$DETECT\_THRESH$, the detection threshold above the background. We 
performed tests to select these parameters, ensuring that we recovered 
all obvious galaxies in the field while not producing large numbers of 
spurious detections. We  chose the rather conservative limits of 
$DETECT\_MINAREA=5$ and $DETECT\_THRESH=1.5$ 
(which provides a nominal $S/N=3.35$), because we think that given the limited 
scientific information to be extracted from the sources close to the 
detection limit, it is better to 
avoid adulterating our catalogs with large numbers of false sources. 
Figure \ref{SE} shows the results of a $SExtractor$ run in a portion 
of the VV 29
field. To estimate the number of isolated spurious detections, we
subtracted the mean sky and changed the sign of all pixels in our images and 
ran $SExtractor$ on them with the same parameters and configuration 
as described 
in Appendix B. The number of spurious detections for \( m_{AB}<28 \)
is very small in all filters, as we can see in Figure \ref{spurious}, but 
have nevertheless been corrected for in the final number counts results.

\subsubsection{Object deblending}\label{deblending}

 Two other very important parameters which govern the deblending process are 
{\it DEBLEND\_NTHRESH} and {\it DEBLEND\_MINCONT}. Typical
values used by other authors are respectively $32$ and $0.01-0.03$. 
At least two teams using HDFN data 
performed part of the detection/deblending process with manual
intervention (Casertano et al. 2000, Vanzella et al. 2001). This
approach may be valid for an isolated field, but we think that it
should not be applied to a large set of observations such as the ones
which will be produced by the ACS GTO program. Not only does manual 
intervention require 
considerable effort, it also introduces subjective biases and 
possible inconsistencies that
make repeatability by other groups difficult, as well as complicating 
quantification of the errors with simulations. Consequently, we decided to look
for values of the deblending parameters able to perform the best
possible 'blind' detection while simultaneously keeping an eye on the
biases introduced by this approach.

\begin{figure}[ht]
\epsfig{file=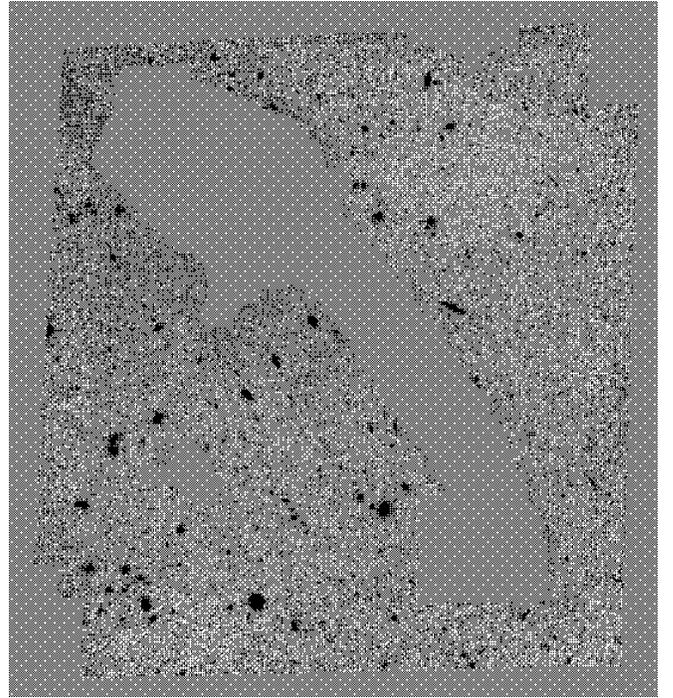,width=\linewidth}
\caption{Detection image for the VV 29 field obtained by combining the
F475W, F606W and F814W filters weighting by the inverse of the
variance, after masking the areas near bright objects where
$SExtractor$ does not perform adequately (see text in Sec. \ref{deblending}).
\label{Tadmask}}
\end{figure}

\begin{figure}[ht]
\epsfig{file=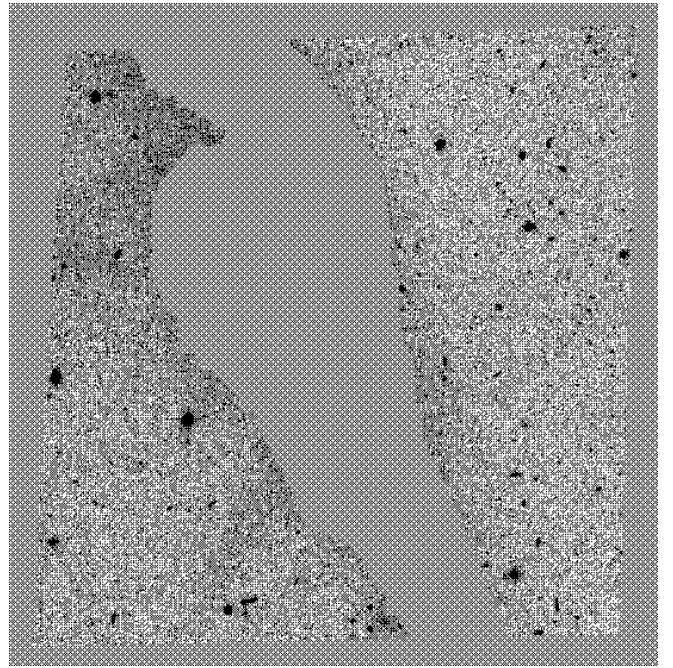,width=\linewidth}
\caption{ Same as Fig. \ref{Tadmask} but for the NGC 4676 field.\label{Micemask}}

\end{figure}

It soon became clear that if we wanted to avoid excessive splitting of
spiral galaxies we would lose some of the faint objects close to the
very brightest galaxies and stars. Because VV~29 is an exceptional field 
with a bright galaxy and associated 
tidal tail spanning the field,  we decided to sacrifice the brightest 
galaxies by
using the automatic procedure described below.  The procedure generates 
masks which enclose the areas around bright objects in which $SExtractor$ 
does not work properly (see Figures {\ref{Tadmask}} and {\ref{Micemask}}). 
Thus, 
we were able to focus on the correct deblending of galaxies in the field.

We first generated a simulated field with the same filters, depth, and
other characteristics as our ACS observations. This was done using
$BUCS$, with galaxy templates extracted from the VV 29 field itself,
guaranteeing the same kind of fragmentation problems encountered in
the real images. This simulated field is processed in exactly the same
way as the real images, and the output $SExtractor$ catalog is matched
and compared with the input catalog. A surprising result was that
about \( \sim 7\% \) of the input objects with \( I<27 \) were not
present in the output catalog, in the sense that no object was
detected within 0.2 arcsec of their positions. This number changed
little for reasonable values of the deblending parameters.  At the
same time, the proportion of spurious objects (present in the output
catalog but not in the input one, and formed from fragments of
brighter galaxies) depended strongly on the deblending parameters,
varying from $ 5\%$ to $ 17\%$.  Since no choice of parameters was
able to totally eliminate both effects at the same time, we decided to
try to make them cancel each other out.

We ran an optimization process on the {\it DEBLEND\_NTHRESH} and {\it
DEBLEND\_MINCONT} parameters by minimizing the difference between the
magnitude distribution of spurious and ``undetected'' objects,
\(D^2_{su}= \Sigma [n_{s}(m)-n_{u}(m)]^{2}\).  This was achieved for
values of 16 and 0.025 respectively.

\begin{figure}[ht]
\epsfig{file=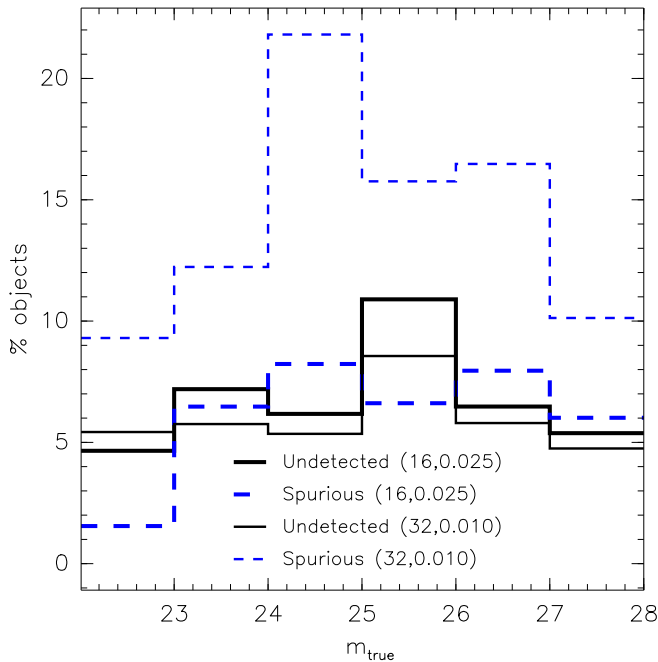,width=\linewidth}
\caption{ Dependence of the number of spurious and undetected objects due
to deblending errors as a function of magnitude for the optimal values 
of the $SExtractor$ parameters $DEBLEND\_MINCONT=0.025$ and $DEBLEND\_NTHRESH=16$ 
(see text). We also show how using different values, e.g. 
$DEBLEND\_MINCONT=0.01$ and $DEBLEND\_NTHRESH=32$ leaves almost unaffected the 
number of undetected galaxies, but more than doubles the number of 
spurious objects, significantly changing the normalization and 
shape of the number count distribution.\label{deblend}}

\end{figure}

By definition, having $D_{su}\approx 0$ conserves the shape and normalization
of the number counts. Visual inspection of the $SExtractor$ aperture
maps shows that most of the parameters in this set produced very good
results (see Figure \ref{SE}).  It should be noted that a significant
part of the differences in the HDF number counts among different
groups (see e.g. Vanzella 2002) may be caused by the deblending
procedure. Changing \emph{DEBLEND\_MINCONT} from 0.025 to 0.01 and 
\emph{DEBLEND\_NTHRESH} from 16 to 32 more than doubles the numbers of
spurious objects, increasing the faint number counts by $\approx 10\%$ 
(Figure \ref{deblend}). The values chosen by Casertano et al. 2002, 32
and 0.03 produce results on the ACS images quite close to the ones 
obtained with our optimal parameter set. 

The parameter set chosen by us ensures that the number and magnitude
distribution of galaxies in the input and output catalogs are the
same. Since the objects used in the simulation are drawn from a real
sample of galaxies, we hope that this will also hold for the
observations.  Now we turn to a delicate question, how to measure the
light emitted by these galaxies.

\subsubsection{Measuring magnitudes}
\label{apertures}

$SExtractor$ provides a plethora of magnitude measurements. Among them
are \emph{MAG\_ISO}, the isophotal magnitude which measures the
integrated light above a certain threshold, \emph{MAG\_AUTO}, an
aperture magnitude measured within an elliptical aperture adapted to
the shape of the object and with a width of k times the isophotal
radius and \emph{MAG\_APER}, a set of circular aperture
magnitudes. The most commonly used magnitude for faint galaxy studies
is \emph{MAG\_AUTO}, with purported accuracies of a few percent for
objects detected at high signal to noise.  However, BFM recently 
claimed that this
measurement technique can be off by more than a magnitude near the
detection limit.

\begin{figure}[ht]
\epsfig{file=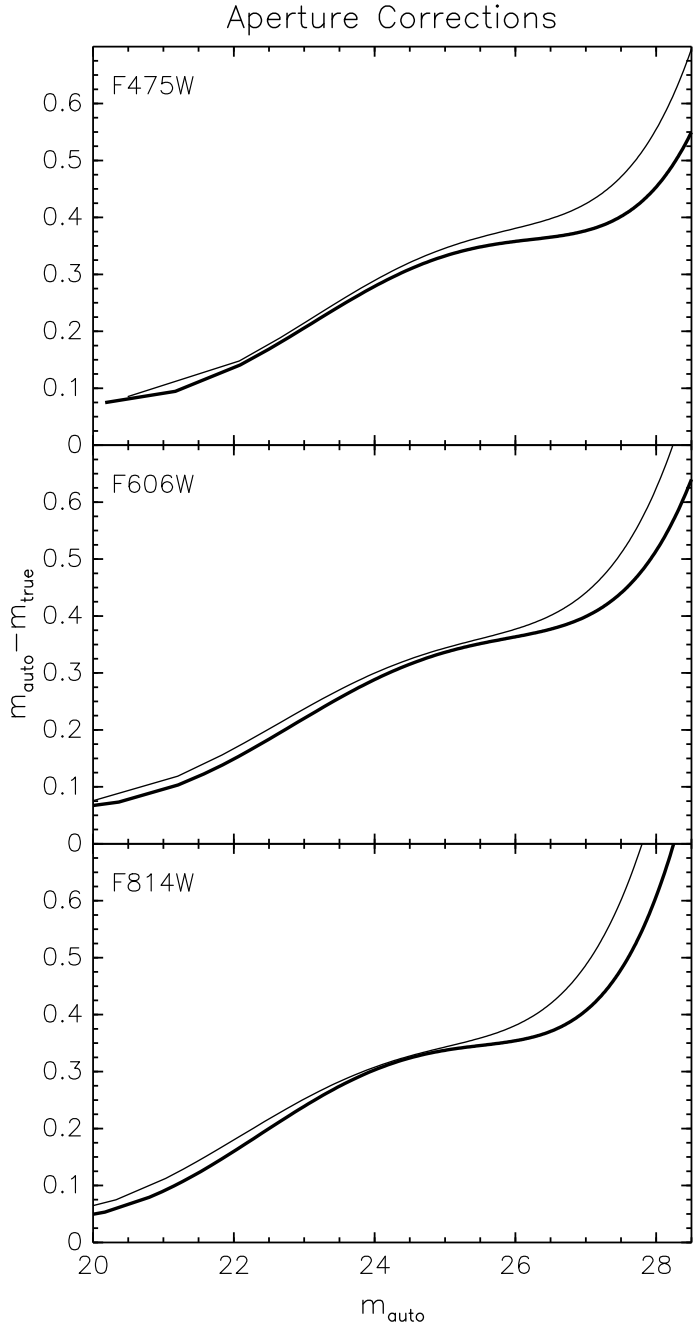}
\caption{Aperture corrections, defined as the difference between the total 'intrinsic' 
magnitudes and the $SExtractor$ $MAG\_AUTO$ magnitude. 
The thick line corresponds to the 
central part of the VV 29 field (see Table 1.), and the thin line to the combination of 
the outer VV 29 field and the NGC 4676 field. Note that the behavior is similar in the 
relatively high S/N regime, but quickly differs at faint magnitudes. \label{fig_apertures}}
\end{figure}

Observational biases in faint galaxy detection and photometry hinder
comparison of distant galaxy samples with lower redshift ones such as
the SDSS (Yasuda et al. 2001, Blanton et al. 2003) and the 2dF (Magdwick
et al. 2002).  To estimate these biases we again resort to simulations
performed with {\it BUCS}. Using the final VV 29 and NGC 4676 images, we
{}``sprinkle'' galaxies with the same redshift, colors, and magnitude
distribution as the objects present in the HDFN onto both fields. We
use HDFN templates instead of VV 29 since the former have much better
color and redshift information, and therefore allow us to recreate
with better accuracy realistic galaxy fields; to avoid overcrowding we
use a surface density of only \( 20\% \) of the observed surface
density. Finally we analyze the simulated images in the same way as
the real images. Because we are interested in comparing the recovered
or ``observed'' magnitudes with the ``true'' ones, we create galaxies
with analytical profiles that have a distribution as similar as
possible to the HDFN real galaxies. We repeat this procedure until \(
\sim 10000 \) galaxies have been added to the NGC 4676 and
VV 29 fields. As expected, we confirm that \emph{MAG\_AUTO} estimates
total magnitudes much better than \emph{MAG\_ISO} or \emph{MAG\_APER}
for reasonable values of the apertures, but there is still a
significant amount of light being left out.  We fit 5-order
polynomials to the median filtered \( m_{auto}-m_{true} \) vs \(
m_{auto} \) data. The results are shown in Figure \ref{fig_apertures}.
We see that our corrections do not rise as dramatically with magnitude
as those of BFM, perhaps
because we are using quite conservative parameters for
\emph{MAG\_AUTO}, an aperture of \( 2.5 \) times the isophotal
radius, and a minimal radial aperture of 0.16 arcsec for faint
objects.  Nevertheless, there is an actual overall dependence on the
depth, especially at very faint magnitudes, where the corrections for
the 'shallow' VV 29 field and the NGC 4676 field are systematically larger
than that of the 'deep' VV 29. The dependence of the correction on
magnitude is quite similar for all filters, and we do observe a
'pedestal' effect which affects even objects with \( m\sim 20 \). In
all filters, the correction increases rapidly when approaching the
detection limit of the field, so one has to be very careful in drawing
conclusions about derived quantities like the luminosity function when
using data close to the detection magnitude limit.

\subsubsection{Color estimation}

\begin{figure}[ht]

\epsfig{file=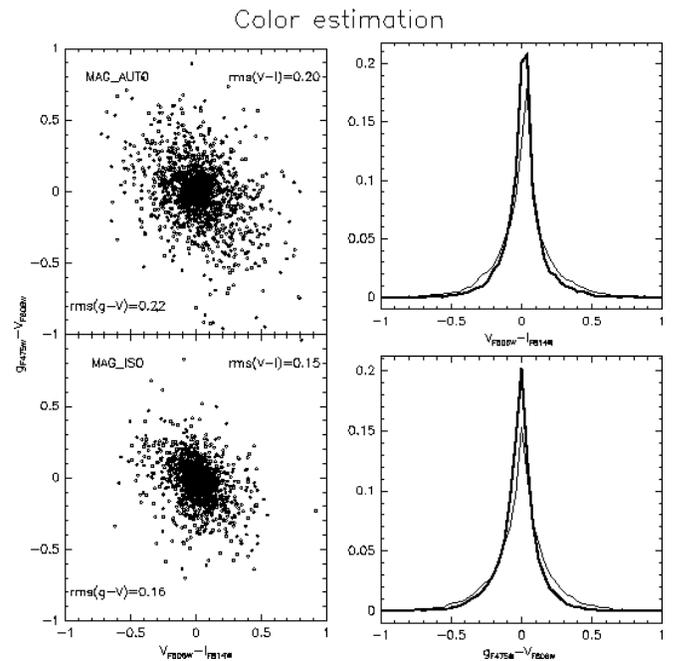,width=\linewidth}
\caption{
On the left plots, the X-axis shows the difference between the $V-I$ 
color measured by SExtractor, and the intrinsic or 'true' colors of the 
galaxies. The Y-axis shows the same quantity, but for the $g-V$ color. 
The bottom left plot corresponds to isophotal and the top left plot 
to Kron magnitudes. Only $20\%$ of the points are plotted for clarity.
Note that the scatter is considerably smaller for isophotal magnitudes. 
The right plots display the color distributions as histograms. The top 
right plot for the $V-I$ colors and the bottom right plot for the 
$g-V$ ones. The thick line corresponds to the isophotal colors, the 
thin line to the Kron ones. 
\label{BestCol}
}
\end{figure}

Accurately measuring the colors of a galaxy is often a different
problem than measuring its total magnitude. In our case, where all
filters have very similar PSFs, using a single aperture defined by the
detection image guarantees that magnitude measurements in all filters
will be affected by the same systematic errors which cancel out when
subtracting the magnitudes to calculate the colors. We again tested
several of the $SExtractor$ measurements and concluded that colors
based on \emph{MAG\_ISO} provide the best estimate of a galaxy's
``true'' colors (provided, of course, that
the object colors \emph{inside} the isophotal threshold are similar to
those \emph{outside} of it).  There seem to be two reasons for this;
first, using an isophotal aperture is more efficient, in terms of
signal--to--noise, than $MAG\_AUTO$, which integrates the light
distribution over regions where the noise is dominant. Second,
although $SExtractor$ tries to correct its aperture magnitude
measurements for the presence of nearby objects, it does not always do so
successfully, and there are a significant number of cases where
the magnitudes are strongly contaminated by the light from close
companions. Isophotal magnitudes are largely free of this problem. The
comparisons between \emph{MAG\_ISO} and \emph{MAG\_AUTO} are shown in
Figure \ref{BestCol}. For bright, compact objects a small aperture
with a diameter of 0.15 arcsec works slightly better than
\emph{MAG\_ISO,} but its performance is equal or slightly worse for
fainter objects, so we decided to use \emph{MAG\_ISO} for all objects.

As expected, the advantages of isophotal magnitudes for estimating
colors also are evident in the photometric redshifts. On average, we
can estimate reliable photometric redshifts for $11\%$ more objects if we use
\emph{MAG\_ISO} instead of \emph{MAG\_AUTO} (see Sec. \ref{photoz}).

\begin{figure}[ht]
\epsfig{file=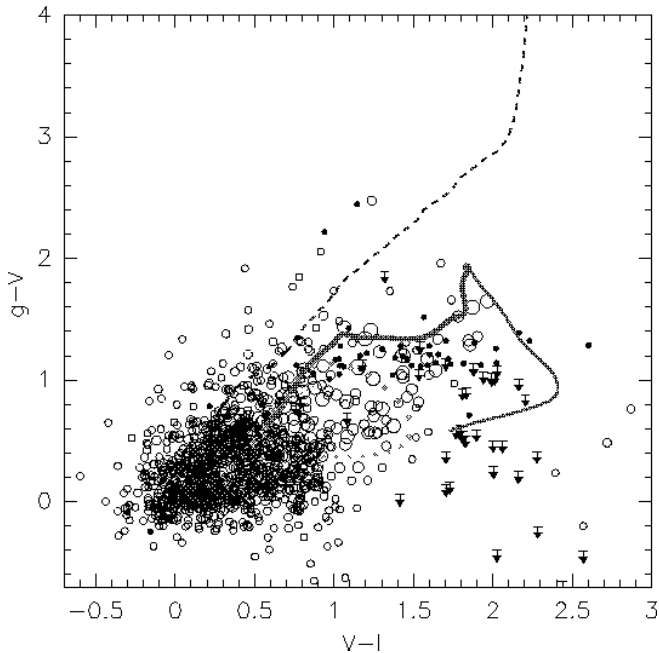,width=\linewidth}
\caption{Color-color plot in the VV 29 and NGC 4676 fields. 
Filled circles represent objects with star class $\ge 0.94$ (i.e. likely to be stars), arrows correspond to objects not detected in the $F475W$ filter 
(we show their $2\sigma$ detection limits), and open circles represent 
galaxies, with the size proportional to their magnitude (the largest 
ones are $I\approx 18$).  
The solid line corresponds to the $z<2$ color track of an 
elliptical galaxy, the dotted line to an Sbc with $z<3$ and the dashed 
line to a starburst. The thicknesses of the curves decrease with redshift. 
For clarity, we only plot $50\%$ of the galaxies.  
\label{ColorColor}}
\end{figure}

 We show color-color plots, together with the tracks corresponding to some 
of the templates introduced below, in Figure \ref{ColorColor}.

\subsubsection{Completeness corrections}\label{masks}

We previously noted that $SExtractor$ is not designed to work near
very bright objects and in general will produce numerous spurious
detections while missing obvious real objects. We are experimenting
with a wavelet-based method to fit the background that may alleviate
the need to perform such steps in the future (White et al., in
preparation). However, at present we must work around this problem in
order to avoid significantly biasing our estimation of the number
count distribution. We masked out areas around bright objects by using
an automatic procedure. First we ran $SExtractor$ and identified only
those objects with areas larger than 20,000 pixels, which are the ones
that typically cause problems with deblending.  The mask was created 
by setting all pixels outside these object to one, and all interior 
pixels to zero. 
To create a ``buffer zone'' around these objects, we convolved the 
mask with a 15 pixel
boxcar filter. In the case of VV 29 we additionally masked a
small area by hand which contained obvious contamination from star
clusters belonging to VV 29 itself.  The final areas that remained after
applying the masking are shown in Figures \ref{Tadmask} and
\ref{Micemask}. The objects in the masked areas are included in the
catalog, but are flagged to show that they are in a masked area where
$SExtractor$ is likely to produce incomplete results.

\begin{figure}[ht]
\epsfig{file=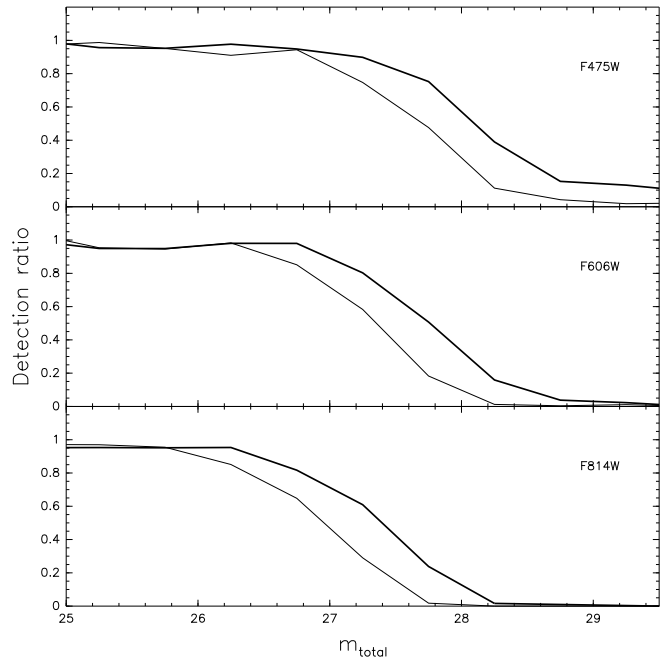,width=\linewidth}
\caption{ Fraction of galaxies detected for each of the filters versus total 
magnitude. The thick line
corresponds to the central region of the VV 29 field, and the thin line
corresponds to the outer region of the VV 29 field plus the NGC 4676
field. Note that the X-axis corresponds to the 'total' intrinsic 
magnitudes\label{ratio}}
\end{figure}

One additional problem is that $SExtractor$'s probability of detecting
an object depends not only on its magnitude, but also on its size,
surface brightness, and other parameters. To measure the
incompleteness as a function of magnitude, we again used the
simulations described in Sec. \ref{deblending}.  We confirmed that the
masking procedure has adequately excluded all the areas where the
galaxy detection is compromised by bright objects. The resulting
detection efficiency is shown in Figure \ref{ratio}.

\subsection{Photometric redshifts}\label{photoz}

The deep, multi-color HDFN observations provided a strong impetus
for developing and using photometric techniques to estimate the
redshifts of faint galaxies. The relatively large number of galaxies
that now have ground-based redshifts provide a benchmark for testing
different photometric redshift methods.  The most widely used HDFN
photometric catalog is that of Fern\'andez-Soto et al. 1999, which
includes PSF matched photometry of the WFPC2 $UBVI$ and ground based
$JHK$ bands.

  Photometric redshift techniques can be broadly divided 
into those which use a library of spectral energy distributions 
(SED), and the ``empirical'' methods, which try 
to model the color-redshift manifold in a non-parametric way 
(see Ben\'\i tez 2000, Csabai et al. 2003 for a detailed discussion). 
The latter require abundant spectroscopic redshifts, and are 
therefore more appropriate for low redshift samples such as the Sloan Digital 
Sky Survey (Csabai et al. 2003). 

\begin{figure}[ht]
\epsfig{file=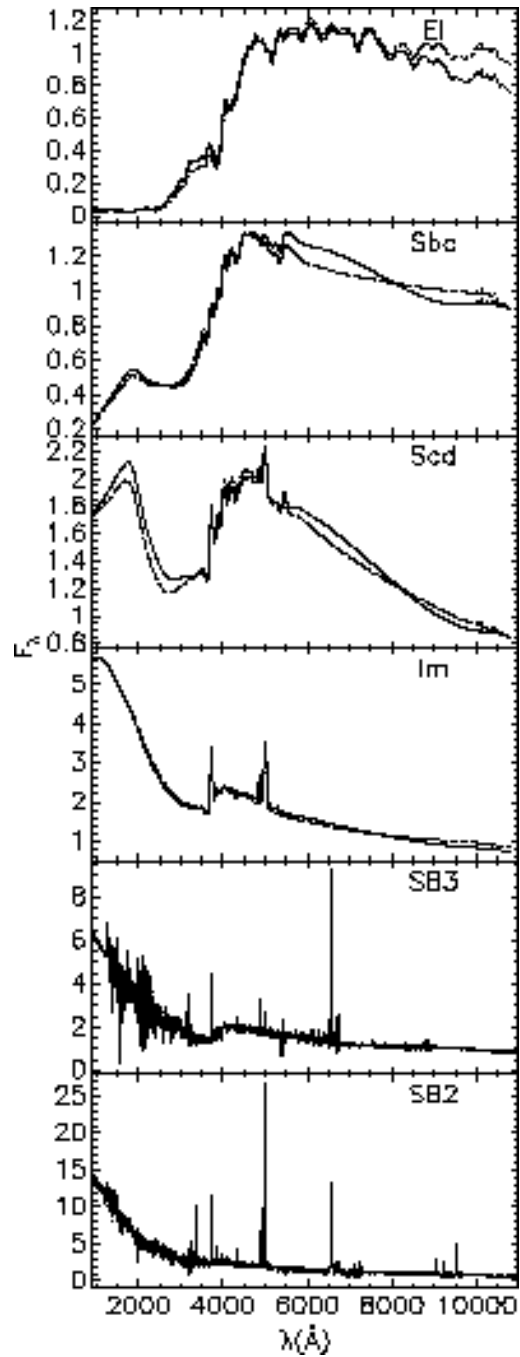}
\caption{The thick lines show ``calibrated'' templates (see text), 
used for photometric redshifts. 
The thin lines correspond to the CWW and Kinney et al. 1996 original 
templates.\label{templates} }
\end{figure}

For faint galaxy samples like the ones presented here, and in general
for ACS observations, the only options are SED-based
techniques, for which a critical issue is the choice of the template
library. At first sight it may seem most logical to use synthetic
galaxy evolution models, like the Bruzual \& Charlot (1993) ones (used
in Hyper-z, Bolzonella, Miralles \& Pell\'o 2000) or the Fioc \&
Rocca-Volmerange (1997) ones, since they take into account age
effects, dust extinction, etc.  However, simple tests quickly showed
that the most effective SEDs for photometric redshift estimation are
obtained from observations of real galaxies, e.g. a subset of the
Coleman, Wu and Weedman (1980) (CWW) spectra augmented with two Kinney
et al.  (1996) starbursts ( Ben\'\i tez 2000, Csabai et al. 2003,
Mobasher et al. in preparation).  But even these templates have
shortcomings.  A detailed comparison of the colors predicted by the
CWW+SB templates and those of real galaxies in several spectroscopic
catalogs show small but significant differences. We
developed a method to trace these differences back to the original
templates, and model them using Chebyshev polynomials, generating a
new set of ``calibrated'' templates which produce much better results
in independent samples. We show these new templates, together with the
original extended CWW set in Figure \ref{templates}.  A detailed
description of this technique will be given elsewhere.

Many observers assume that measuring accurate photometric redshifts
requires as many as five or more filters. However, as we show here,
useful redshift information can be derived from as few as three
filters by using a Bayesian approach. The problem of determining
photometric redshifts can be stated as (Ben\'\i tez 2000):
\begin{equation}
\label{bas}
p(z|C,m_{0})=\sum _{T}p(z,T|C,m_{0})\propto \sum
_{T}p(z,T|m_{0})p(C|z,T)
\end{equation}
where $z,C$ and $m_0$ are respectively the redshift, colors and
magnitude of a galaxy, and $T$ corresponds to the templates, or 
spectral energy distributions (SED) used to
estimate its colors. The term $p(C|z,T)$ is the likelihood, and the
differences between Bayesian photometric redshifts and
maximum-likelihood or $\chi^2 $ ones arise from the presence of the prior
$p(z,T|m_{0})$ and the marginalization over all the template types.
The redshift-type prior is neither more nor less than the expected
redshift distribution for galaxies of a given spectral type as a
function of magnitude. It contains what we know about a galaxy's
redshift and type just by looking at its magnitude. In most cases this
is very little, of course, but it is obvious that using this
information is just translating common sense to a mathematical form:
brighter galaxies tend to be at lower redshifts than fainter ones. 

\begin{figure}[ht]
\epsfig{file=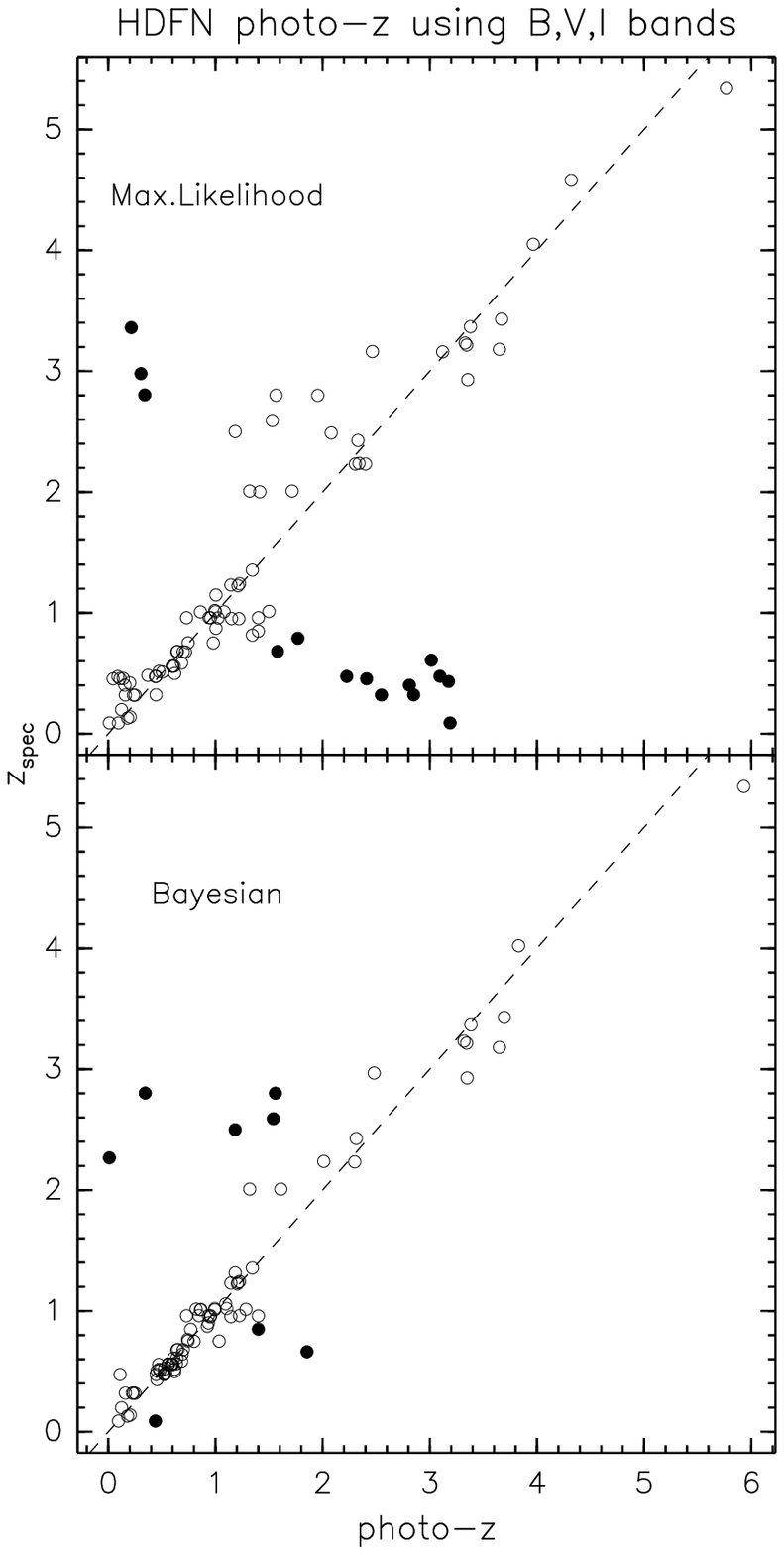}
\caption{Comparison between spectroscopic and photometric redshifts 
obtained using, using 
basically the same filter set as the VV 29 and NGC 4676 observations. 
Note that the axes are inverted with respect to the typical orientation  
in this kind of plot. This is done to better show the effects of using 
photometric redshifts to select galaxy samples. In the case 
of Maximum Likelihood photometric redshifts, almost one third of the $z>1.5$ objects would 
be misclassified low redshift objects, biasing high any estimate of the 
the luminosity function, star formation rate, etc. obtained with them. 
The dashed line has slope 1 and it does not represent a fit to the data. 
The filled circles correspond to the objects classified as outliers 
in Table 3. Note that since the axis are inverted, the outliers were 
selected based on their {\it horizontal} distance to the dashed line 
with respect to the rms fluctuation. 
 \label{bpz}}
\end{figure}

There is a persistent prejudice that using a prior 
will ``bias'' the redshift estimate, making the data unfit for various
scientific applications like measurement of the luminosity function,
whereas maximum likelihood estimates are free of such problems. It is
easy to show that this is completely unjustified from the point of
view of probability theory. It is clear that using 
maximum likelihood is similar, in this particular setting, to using a
``flat'' prior, i.e., $p(z,T|m_0)=const$.  This means that using
maximum likelihood (or equivalently $\chi^2$) is not assumption-free; 
on the contrary, it is
similar to assuming that the redshift distribution of galaxies is 
{\it flat at all magnitudes}. To obtain
such an observed redshift distribution one has to contrive a
luminosity function with enormous evolution rates, therefore tending
to significantly ``overproduce'' the number of high-z galaxies.  This
is clearly shown in our tests below (see Figure \ref{bpz}).

\begin{deluxetable*}{lcccc}
\tabletypesize{\scriptsize} 
\tablecaption{Performance of Bayesian and Maximum Likelihood photometric redshift methods} 
\tablewidth{0pt} 
\tablehead{
\colhead{sample}& 
\colhead{\(n_{gal}\)}& 
\colhead{mean}& 
\colhead{rms\tablenotemark{a}}&
\colhead{\( n_{out} \)}}
\startdata 
$z<1.5$, bayesian       & 73 & -0.002 & 0.073 & 5  \cr 
$z<1.5$, max.likelihood & 58 & -0.018 & 0.133 & 3 \cr 
$z>1.5$, bayesian       & 16 &  0.004 & 0.081 & 3  \cr 
$z>1.5$, max.likelihood & 31 & -0.04  & 0.126 & 11 
\enddata
\tablenotetext{a}{The root-mean-squared has been calculated after eliminating the most obvious outliers ($n_{out}$) by sigma-clipping}

\end{deluxetable*}

 Ideally one would want to use the 'real' redshift distribution
$p(z,T|m_0)$ of the field as the prior, but this is usually unknown,
since it is the quantity we want to measure. But it is clear that an
analytical fit to the redshift histogram from a similar blank field,
like the HDFN, is always a much better approximation---in spite of the
cosmic variance---than a flat redshift distribution. Thus using
empirical priors such as the ones introduced in Ben\'\i tez (2000),
does in fact considerably reduce the biases introduced by
maximum-likelihood methods. Simple comparisons like the one below using 
the same dataset and
template sets show that, as expected, Bayesian probability gives
consistently more accurate and reliable results than
maximum-likelihood or $chi^2$ techniques (see also Ben\'\i tez 2000, Csabai et
al. 2003, Mobasher et al. in preparation).

To test how well we can expect to estimate photometric redshifts with 
our data, we
performed the following test. We ran BPZ with the same set of
parameters described in Appendix C, but using only the WFPC2 BVI
photometry from the Fern\'andez-Soto et al. 1999 catalog. This is
almost identical in filter coverage and depth to the observations
discussed here, and therefore serves as an excellent test of the
performance of our photometric redshifts. The results, both for the 
Bayesian ($z_B$)
and maximum likelihood photometric redshifts ($z_{ML}$) are shown
in Figure \ref{bpz} and Table 3. In the lower plot we excluded
those objects with Bayesian odds \(O <0.9 \) (Ben\'\i tez 2000), about
\( 1/3 \) of the sample, and performed a similar
preselection for $z_{ML}$ by excluding the objects with the highest
values of \( \chi ^{2} \), up to \( 1/3 \) of the total. We see that
despite this pruning of the data, the number of ``catastrophic'' maximum 
likelihood outliers (error $\ge 3 \sigma$) 
seriously affects any scientific analysis, especially in the $z>1.5$ range, 
where $1/3$ of the objects selected using Maximum Likelihood photometric 
redshifts happen 
to be low-redshift galaxies. This is a good example of the 
tendency to overproduce the number of high$-z$ galaxies of the 
maximum likelihood or $chi^2$ methods discussed above. 

\begin{figure}[ht]
\epsfig{file=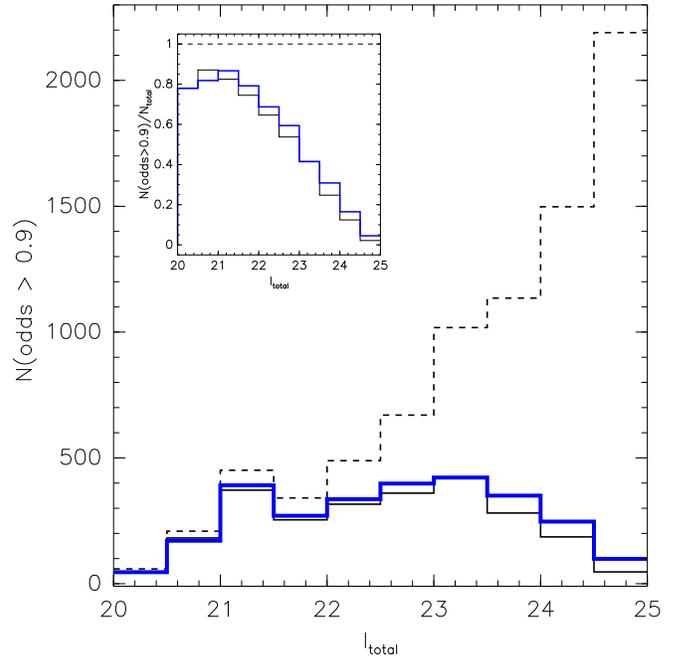,width=\linewidth}
\caption{Distribution of galaxies with Bayesian $odds > 0.9$ (photometric redshifts 
with lower values are  
unreliable) in our 
simulations as a function of magnitude. This shows that there are few objects with good quality photometric redshifts 
for $I\gtrsim 24.5$. The thick line corresponds to isophotal magnitudes, 
and the thin line to Kron apertures; the figure shows that isophotal 
magnitudes improve the accuracy of the photometric redshifts. The dashed line 
corresponds to the total number of objects.
\label{figphotoz1}}
\end{figure}

\begin{figure}[ht]
\epsfig{file=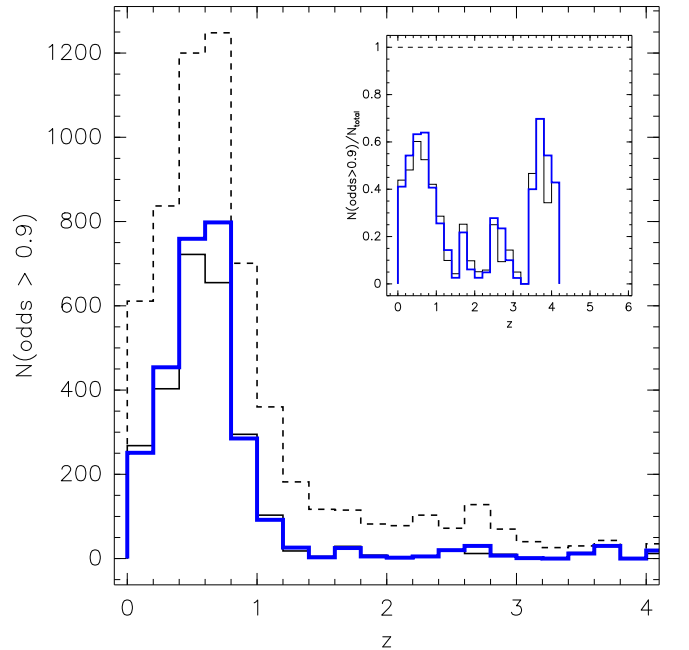,width=\linewidth}
\caption{Redshift distribution of $I<25$ galaxies. The meaning of the 
different types of lines are the same as in the previous figure. 
\label{figphotoz2}} 
\end{figure}

\begin{figure}[ht]
\epsfig{file=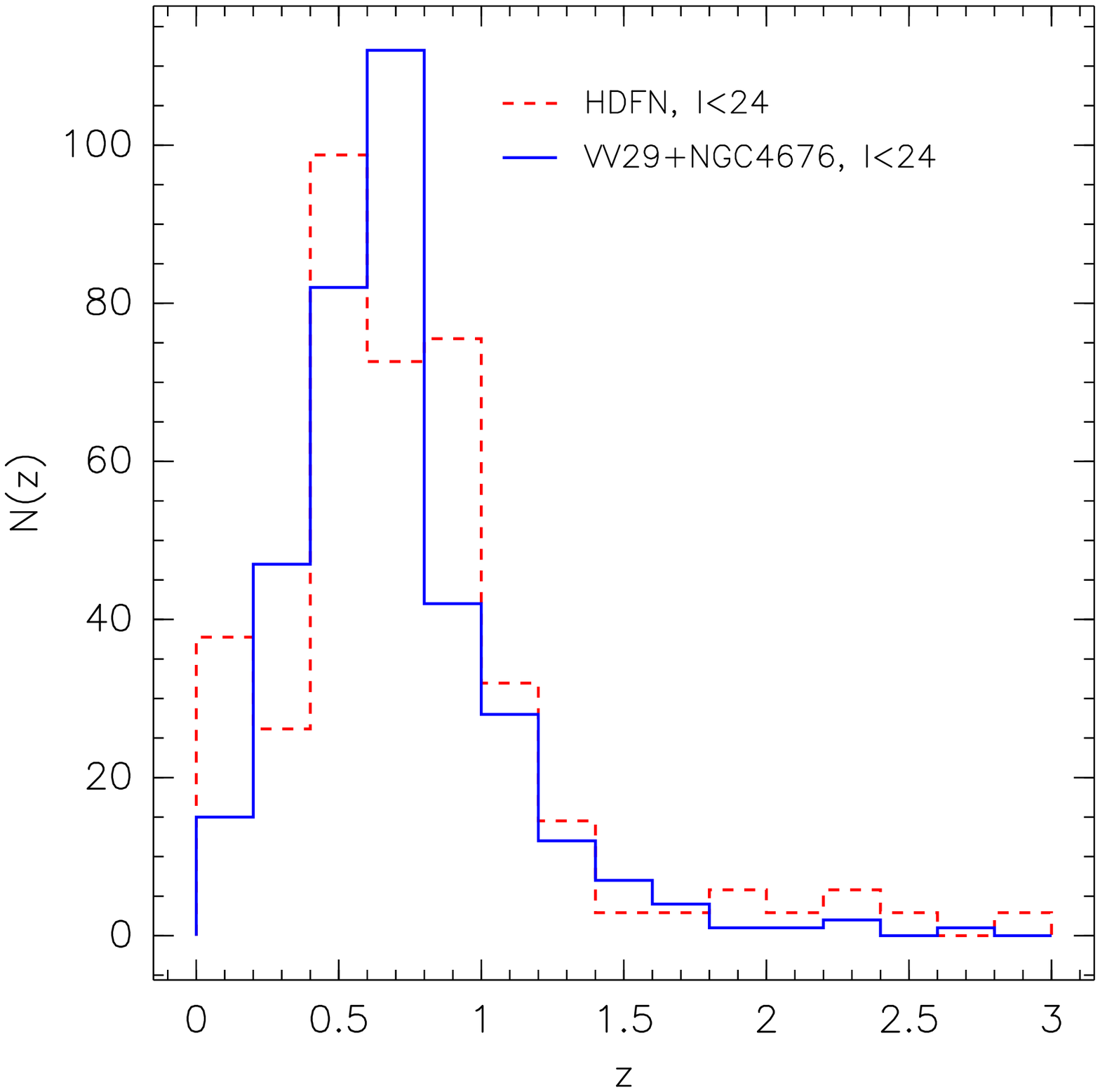,width=\linewidth}
\caption{Comparison between the redshift distribution of all the galaxies with 
$I<24$ in our observations (solid line) and the HDFN (dashed line) scaled 
to our total effective area. 
\label{figphotoz3}}
\end{figure}

We also performed a test based on the simulations described above to
determine the ``efficiency'' of our photometric redshifts as a function of
magnitude and redshift. We looked at the number of galaxies with
Bayesian odds $O>0.9$ in the output of the simulations described in
the section above as a function of {\it total} magnitude in the
$F814W$ band, $I_{total}$, and 'true' redshift. The results are shown
in Figures \ref{figphotoz1} and \ref{figphotoz2}.  They show that, using this
limited filter set, we can only expect to estimate reliable photometric 
redshifts for bright, $I\lesssim 24$ objects, and only for certain regions of
the redshift range. This should be taken into account when using the
photometric redshifts in the catalog. Figure \ref{figphotoz3} compares 
the redshift distribution in our fields with that of the HDFN for 
galaxies with $I_{F814W}<24$.

Tables 6 and 7 give the  photometric and positional information for 
those objects whose
photometric redshifts have very high values of the Bayesian odds, and
which therefore can be expected to be quite accurate and reliable.
Based on previous experience and comparison with other catalogs such
as the HDFN, we expect an rms accuracy of $\approx 0.1 (1+z)$ and only 
a few percent of objects with ``catastrophic'' redshift errors.  We also 
provide photometric redshift information
for the rest of the objects in both fields as explained below, but we
note here that, as BPZ indicates, their redshifts are much more
uncertain.

\subsection{The catalogs}

 For each of the fields observed by ACS we provide two catalogs which will be published electronically and also 
made available at the ACS website (http://acs.pha.jhu.edu).

\subsubsection{Photometric Catalog}
These catalogs contain all the objects detected by $SExtractor$ in each of the fields. 
We decided not to purge the spurious detections, but effectively eliminated them by selecting only objects with $mask\_flag=0$. The photometric catalog contains the following columns:

\noindent - $ ID$. This is the $SExtractor$ ID number in the output catalog. 

\noindent - $ RA, DEC$. These are the right ascension and declination, calibrated with the Guide Star Catalog II. They have relative accuracies 
of $\lesssim 0\farcs1$. 

\noindent - $ X,Y.$ Pixel coordinates in the images.

\noindent - $ g_{auto}, V_{auto}, I_{auto}$. 
These are $SExtractor$, {\it uncorrected} Kron elliptical magnitudes $MAG\_AUTO$. 
They should be used as the best estimate of the total magnitude of a galaxy, 
although they miss an increasingly large fraction of the light at faint magnitudes. 
But, as argued above, for color estimation or photometric redshifts we 
recommend isophotal magnitudes.

\noindent - $ g_{iso}, V_{iso}, I_{iso}$. These are $SExtractor$ isophotal ($MAG\_ISO$) magnitudes. 

\noindent - $ FWHM$. Full width at half maximum in {\it pixels},
recalling that the scale of the images is $0\farcs05$ pixel$^{-1}$.

\noindent - $ star class$. $SExtractor$ star/galaxy classifier. We consider as stars or point sources all objects which have a value of this 
parameter $\geq 0.94$, and set all their photometric redshift parameters to zero.

\noindent - $ flag$. $SExtractor$ detection flag. 

\noindent - $ mask flag$. If the value of this flag is 1, it indicates
that the object is in an area strongly affected by incompleteness or
the presence of spurious objects. For most science uses only objects
with $mask flag=0$ should be selected.

\subsubsection{Photometric redshift Catalog}

We estimated Bayesian photometric redshifts for galaxies in our catalog
using the parameters specified in Appendix C. The main difference
relative to the method presented in Ben\'\i tez 2000 is that we used
the template library described above.  For a more detailed discussion,
see Ben\'\i tez 2000.

\noindent - $z_b$. Bayesian photometric redshift, or maximum of the redshift 
probability distribution. 

\noindent - $z_b^{min},z_b^{max}$.   Lower and upper limits of the redshift 
probability $95\%$ confidence interval. Note that in some cases, this 
probability distribution is multimodal, so these values or $z_b$ are not 
very meaningful. 

\noindent - $t_b$.  Bayesian spectral type. The types are El (1), Sbc(2), 
Scd(3), Im(4), SB3(5), and SB2(6).

\noindent - $odds$.  Bayesian odds. This is the integral of the redshift 
probability distribution in a region of 
$\approx 0.2(1+z_b)$. If close to 1, it means that the redshift 
probability is narrow and has a single peak. Very low values 
of the odds indicate that the color/magnitude information is 
almost useless to estimate the redshift. 

\noindent - $z_{ml}$. Maximum likelihood redshift. We provide this to 
allow users to compare with the value of $z_b$, and also to understand 
the effects of the prior on the redshift estimate. As an estimate of 
the redshift we recommend the use of $z_b$, which has been proved to 
be more accurate and reliable. 

\noindent - $t_{ml}$. Maximum likelihood spectral type. 

\noindent - $\chi^2$. This is the value corresponding to the maximum likelihood redshift/spectral type fit.

\section{Number counts}

\begin{figure}[ht]
\epsfig{file=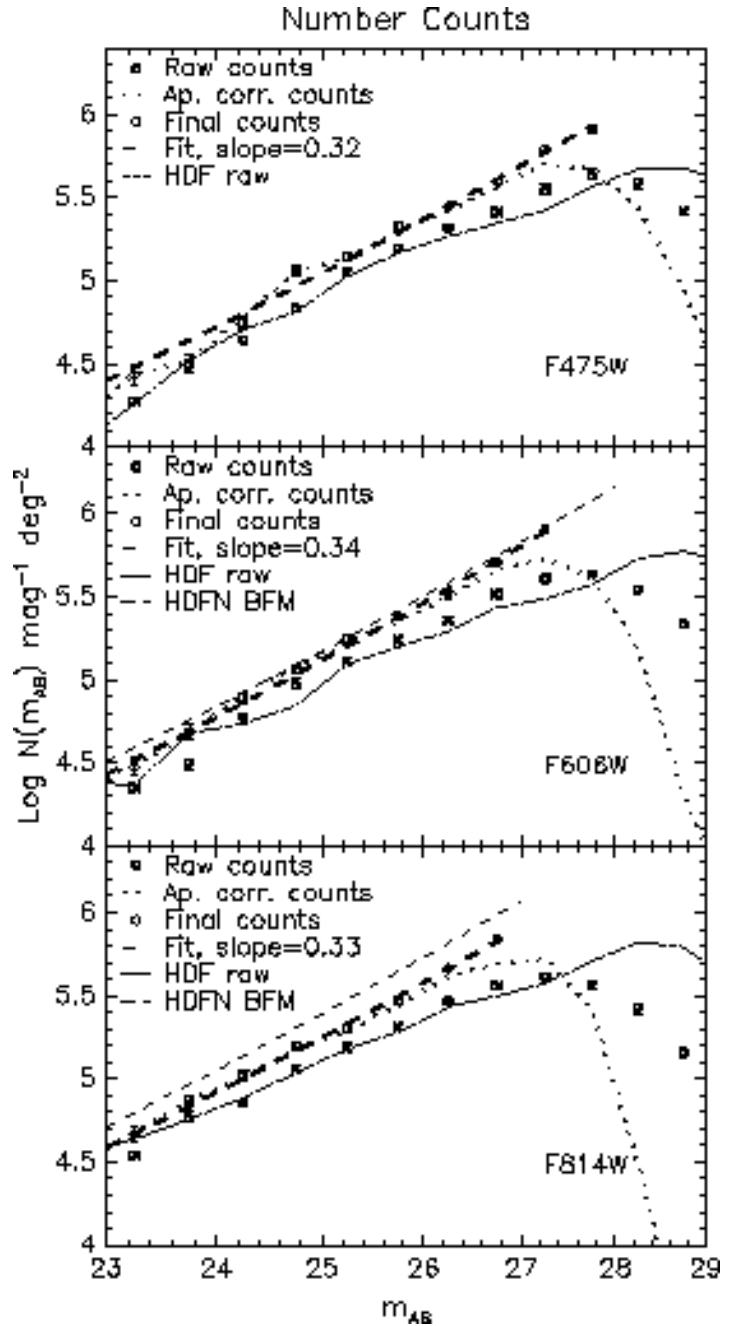}
\caption{Number counts distributions. The filled squares , 
the dotted line is our counts after applying aperture corrections, 
and the open circles show the final number counts after correcting for 
incompleteness. The thick dashed line is a least square fit to the 
open circles. 
The continuous thin lines are the number counts in the HDFS+HDFN  
(Casertano et al. 2000). In the two bottom plots we also show the corrected 
counts from Bernstein, Freedman \& Madore 2002, which have slopes and 
normalizations close to our results even though they were obtained with 
a different approach.
\label{NC}}
\end{figure}

 Our number counts  are shown in Figure \ref{NC} and listed in Table 4. We
plot the raw number counts, as measured by $SExtractor$, the aperture
corrected counts as described in Section \ref{apertures}, and finally 
correct for spurious detections and incompleteness to estimate the final number counts. The
result can be very well fitted by a straight line with slopes $0.32\pm 0.01$,
$0.34\pm 0.01$ and $0.33\pm 0.01$ respectively in the $F475W$, $F606W$ and $F814W$ bands 
(Table 5). This is in excellent agreement with the results of BFM, who measure slopes 
of $0.33\pm 0.01$ and $0.34\pm 0.01$ respectively in the $F606W$ and $F814W$ bands. 
The normalization is in remarkably good agreement too, taking into account the size of the
fields and the very different correction procedures followed. 

 There is a significant difference in the raw number counts between our
data and the HDF's, of about \( 40\% \) in the g and V bands and \(
20\% \) in the I band, in the sense that we find more objects than
Casertano et al. (2000).  Although $1/4$ of this difference goes away
when we use a detection image formed by only our V and I images
(instead of $gVI$), most of the difference is probably due to the fact that 
our PSF is more compact, with significantly less light at large radii than 
for WFPC2.  Therefore the apertures used by $MAG\_AUTO$ will enclose more
of the light from each galaxy, creating a steepening effect similar to
the one produced by the aperture corrections.  To test for possible
contamination from star clusters belonging to the central galaxies, we
mapped the distribution of extended objects with colors similar to
obvious star clusters. Since they are distributed homogeneously
across the field and are not particularly concentrated toward the central
galaxies, they are probably not a large contaminant.

\begin{deluxetable*}{lrrrrrrrrrrr}
\tablecolumns{12}
\tabletypesize{\scriptsize} 
\tablecaption{Number Counts} 
\tablewidth{0pt} 
\tablehead{
\colhead{}   &  \multicolumn{3}{c}{$g_{F475W}$} &   
\colhead{}   &  \multicolumn{3}{c}{$V_{F606W}$} &
\colhead{}   &  \multicolumn{3}{c}{$I_{F814W}$}  \\
\cline{2-4} \cline{6-8} \cline{10-12} \\   
\colhead{$m_{AB}$}  	& 
\colhead{$N$}   		& 
\colhead{$N_{raw}$} 	&
\colhead{$dN_{raw}$} 	& 
\colhead{}			&
\colhead{$N$}        	& 
\colhead{$N_{raw}$}  	& 
\colhead{$dN_{raw}$} 	& 
\colhead{}			&
\colhead{$N$}       	& 
\colhead{$N_{raw}$} 	&
\colhead{$dN_{raw}$} 
}
\startdata 
23.25 & 27177 & 18909 & 3103 && 30072 & 22486 & 3384 && 46166 & 34751 & 4209 \\
23.75 & 31849 & 30152 & 3920 && 48467 & 30663 & 3953 && 73853 & 59282 & 5499 \\
24.25 & 55978 & 43950 & 4734 && 77776 & 58771 & 5475 && 103432 & 71547 & 6041 \\
24.75 & 115058 & 68481 & 5910 && 115580 & 95056 & 6964 && 158452 & 113454 & 7609 \\
25.25 & 140421 & 112943 & 7592 && 177131 & 129297 & 8123 && 202678 & 154339 & 8876 \\
25.75 & 213130 & 154339 & 8876 && 239108 & 172737 & 9390 && 292149 & 208511 & 10317 \\
26.25 & 277901 & 205955 & 10254 && 321145 & 227420 & 10775 && 459216 & 295901 & 12292 \\
26.75 & 390832 & 259105 & 11502 && 508518 & 326054 & 12903 && 689457 & 368472 & 13717 \\
27.25 & 616540 & 353651 & 13438 && 799663 & 405268 & 14386 && 1282821 & 406290 & 14404 \\
\enddata
\tablecomments{Corrected numbers counts $N(m)$ in the three filters.
We also present the raw number counts $N_{raw}(m)$, based on 
the $MAG\_AUTO$ magnitudes provided by SExtractor, and their $\sqrt(N)$ 
errors for comparison. The raw counts are measured on a 
$14.1$ arcmin$^2$ area, but all quantities are normalized to 
1 sq. degree area
}
\end{deluxetable*}

\begin{deluxetable*}{lllll}
\tabletypesize{\scriptsize} 
\tablecaption{Number count slopes} 
\tablewidth{0pt} 
\tablehead{
\colhead{Filter}& 
\colhead{$\alpha$}& 
\colhead{mag. range}&
\colhead{$\alpha_{BFM}$}& 
\colhead{mag. range (BFM)}}
\startdata 
$F475W$ & $0.32\pm 0.01$ & $22<m_{AB}<28$   &     -         & -  \cr 
$F606W$ & $0.34\pm 0.01$ & $22<m_{AB}<27.5$ & $0.33\pm 0.01$  & $22 < m_{AB} < 27$ \cr 
$F814W$ & $0.33\pm 0.01$ & $22<m_{AB}<27$   & $0.34\pm 0.01$  & $22 < m_{AB} < 27$ \cr 
\enddata
\tablecomments{
Slopes for number count fits of the form $N(m)\propto 10^{\alpha m}$, both for 
the results presented in these paper and those of Bernstein, Freedman and Madore (2002a,b). 
We also include the magnitude interval on which the fit was performed.}
\end{deluxetable*}

\begin{figure}[ht]
\epsfig{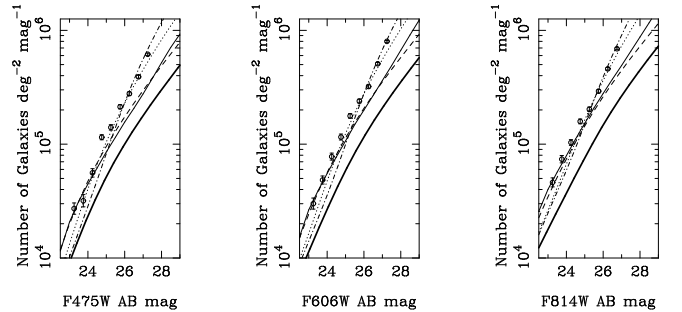}
\caption{Comparison with number count models. 
The points correspond to our final corrected number counts (same 
as the open squares in the previous plot). The thick solid line 
represents a no-evolution model and the dashed line a passive 
luminosity evolution (PLE) model. 
The dotted and dotted-dashed line represent PLE models including merger of 
galaxies following respectively the Guiderdoni \& Rocca-Volmerange (1990) and 
the Broadhurst, Ellis \& Glazebrook (1990) prescriptions. The thin solid 
line represents the LCDM predictions of Nagashima et al. (2002). 
See the text for details about the model generation and comments.
\label{NCM}}
\end{figure}

Galaxy number counts (especially at very faint magnitudes) provide
important constraints on galaxy formation and evolution 
(Gronwall \& Koo 1995). 
We compare our results with some simple galaxy number count 
generated using the public software {\it ncmod} (Gardner 1998) in 
Figure \ref{NCM}.We make use of the recently derived $B$-band luminosity 
function from the COMBO-17 survey (Wolf et al. 2003), derived at $z\sim 0.3$ 
and which features a rather steep slope ($\alpha=-1.5$) for the faint end of 
the luminosity function. We assume a flat, $\Omega_{\Lambda}=0.7$ 
cosmology with $H_0=70$ km s$^{-1}$ Mpc$^{-1}$. We generate K+e 
corrections using GISSEL96 synthetic templates (Bruzual \& Charlot 1993). 
This pure luminosity evolution model works reasonably well for magnitudes 
brighter than $25$ in all bands, but drops significantly below the 
observed counts at fainter magnitudes, falling short by a factor of 
$\sim 2$ by $m \sim 27$. It is clear that our data have reached the magnitude 
levels at which merging (as expected by hierarchical models of galaxy 
formation) is important. Thus, we also plot the predictions of two 
luminosity  evolution models with simple (and rather strong) merging 
prescriptions. Guiderdoni \& Rocca-Volmerange (1990) proposed a merging 
rate of (1+z)$^{\eta}$, with $\eta=1.5$,  
but their predicted counts also fall short of 
our measurements. Following Broadhurst, Ellis \& Glazebrook (1992), 
Glazebrook et al. (1994) proposed a model where the merging rate is 
proportional to $1 + Q \times z$. We use $Q=4$ 
(such that a present day galaxy is the result of a merger between 
$\sim 4$ sub-units at $z \sim 1$.) and find that 
this prescription fits well the counts in the $m_{AB}>25.5$ range, 
producing a distribution which has the measured slopes and 
normalization in all our bands. 

Finally, we also include the hierarchical model predictions 
of Nagashima et al. (2002) which, although include both the luminosity 
evolution and merging (as predicted by hierarchical formation theory) of 
galaxies, also underpredict the observed number counts by a 
factor of $\sim 2$.  

\begin{figure}[ht]
\epsfig{file=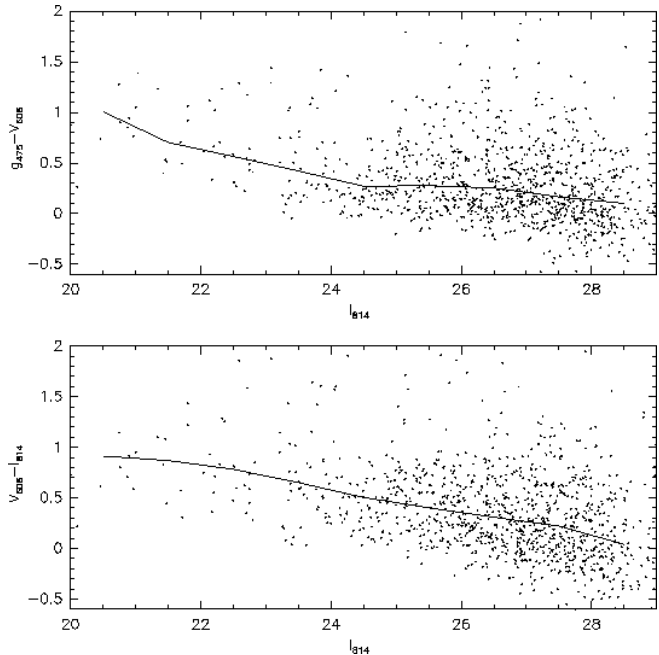,width=\linewidth}
\caption{Median color as a function of magnitude. The upper and lower figures 
show respectively the $g-V$ and $V-I$ isophotal colors of galaxies plotted as 
a function of their $MAG\_AUTO$  $I$ magnitude. The solid lines represent the 
median color of the galaxies.
\label{Colormag}}
\end{figure}

The color distribution of galaxies with magnitude also provides important
constraints on galaxy evolution models.  Figure \ref{Colormag} shows
the observed and median galaxy colors as a function of magnitude.
The typical color of {\it detected} $I\approx 28$ galaxies is $g-V\sim 
V-I\sim 0.15$ are similar to those of blue starbursting galaxies at 
$1.2 \leq z \leq 2.6$, as expected from the results of Ben\'\i tez 
2000 for the redshift distribution of faint galaxies in the HDFN. 

We would like to note that although the prediction of the Glazebrook 
et al. (1994) model fits the data satisfactorily, any result of 
this kind is very sensitive to the choice of luminosity function 
parameters, and that in any case it is very unlikely that the 
evolution of the galaxy population is will be described by such a 
simple scheme. Although it is beyond the scope of this paper to explore 
more sophisticated models of galaxy number counts, we 
hope that these new data and our effort to remove instrumental 
biases and other corrections which hinder the measurement of 
the number count distribution will aid future modeling efforts.

\section{Conclusions}

  We present the analysis of the faint galaxy population in the
Advanced Camera for Surveys (ACS) Early Release Observation fields VV
29 (UGC 10214) and NGC 4676. These were the first science observations 
of galaxy fields with ACS and show its efficiency compared with 
the previous Hubble Space Telescope optical imaging preferred instrument, 
WFPC2. The observations cover a total area of 26.3 arcmin$^2$, with an 
effective area for faint galaxy studies of 14.1 arcmin$^2$,  
and have depths close to that of the Hubble Deep Fields in the central 
and deepest part of the VV 29 image, with $10\sigma$ detection limits for 
point sources of $27.8$, $27.6$ and $27.2$ AB magnitudes in the 
$g_{F475W}$, $V_{F606W}$ and $I_{F814W}$ bands respectively.

The measurement of the faint galaxy number count distribution is still
a somewhat controversial subject, with different groups arriving at
widely varying results even on the same dataset.  Here we attempt to
thoroughly consider all aspects relevant for faint galaxy counting and
photometry, developing methods which are based on public software like
$SExtractor$ and $BUCS$, and therefore easy to reproduce by other 
astronomers.

Using simulations we determine the best $SExtractor$ parameters for
the detection of faint galaxies in deep HST observations, paying
special attention to the issue of deblending, which significantly
affects the normalization and shape of the number count
distribution. We also confirm, as proposed by BFM, that Kron-like 
($MAG\_AUTO$) 
magnitudes, like the ones generated by $SExtractor$, can miss more 
than half of the light of faint galaxies. This dramatic effect, which 
strongly changes the shape of simulated number count distributions, 
depends sensitively not only on the characteristics of the observations, 
but also on the choice of $SExtractor$ parameters, and needs to be taken 
into account to make meaningful comparisons with theoretical models or 
between the results of different authors.

We present catalogs for the VV 29 and NGC 4676 fields with photometry 
in the $F475W, F606W$ and $F814$ bands. We show that combining the 
Bayesian software BPZ with superb ACS data and new templates enables us 
to estimate reliable photometric redshifts for a significant fraction 
of galaxies with as few as three filters. 

After correcting for selection effects we find that the faint number
counts have slopes of $0.32\pm 0.01$ for $22<g_{F475W}<28.$, 
$0.34 \pm 0.01$ for $22<V_{F606W}<27.5$ and $0.33\pm0.01$ for 
$22<I_{F814W}<27$, and do not flatten (except
perhaps in the $F475W$ filter), up to the depth of our observations.
Our results agree well with those that BFM obtained with different 
datasets and techniques ($0.33\pm 0.01$ for $22<V_{F606W}<27$ and 
$0.34\pm0.01$ for $22<I_{F814W}<27.$). This is encouraging, and shows 
that it is possible to perform consistent measurements of galaxy 
number counts if the selection effects are properly taken into account.

While some may argue that any corrections--even well motivated ones
such as those we use here--are model dependent, given the
magnitude of the selection effects, applying some correction is better
than none at all, as the widely varying results on faint galaxy number
counts demonstrate. We have presented a methodology based on freely available 
software that enables a consistent comparison across different datasets 
and against theoretical results.

We compare our counts with some simple traditional number count models 
using the software $ncmod$ (Gardner 1998). At brighter magnitudes 
($m_{AB}<25$) the counts are well approximated by a passive luminosity 
evolution model based on the steep slope ($\alpha=-1.5$) quasi-local 
luminosity function from the COMBO-17 survey. This model underpredicts 
the faint end by a factor $\sim 2$, and it is necessary to introduce 
the merging prescription of Glazebrook et al (1994), $\phi^*\propto (1+Qz)$, 
which with $Q=4$ produces good fits to both the slope and number count 
normalization at $m_{AB}>25.5$ in all our filters.  

\acknowledgements

ACS was developed under NASA contract NAS 5-32865, and this research 
has been supported by NASA grant NAG5-7697. We are grateful for an 
equipment grant from  Sun Microsystems, Inc.  
The {Space Telescope Science
Institute} is operated by AURA Inc., under NASA contract NAS5-26555.
We are grateful to K.~Anderson, J.~McCann, S.~Busching, A.~Framarini, 
S.~Barkhouser,
and T.~Allen for their invaluable contributions to the ACS project at JHU. 
We also thank Masahiro Nagashima for useful comments.

\clearpage

\LongTables

\label{table_VV 29}
\begin{deluxetable}{lrrrrrrrrrr}
\tabletypesize{\scriptsize}
\tablecaption{Photometric Redshift catalog in the VV 29 field}
\tablewidth{0pt} 
\tablehead{
\colhead{ID \tablenotemark{a}}&
\colhead{RA \tablenotemark{b}}&
\colhead{DEC \tablenotemark{c}}&
\colhead{X}&
\colhead{Y}&
\colhead{$z_b$ \tablenotemark{d}}&
\colhead{$g_{F475W}$ \tablenotemark{e}}&
\colhead{$V_{F606W}$ \tablenotemark{f}}&
\colhead{$I_{F814W}$ \tablenotemark{g}}&
\colhead{FWHM \tablenotemark{h}}&
\colhead{s \tablenotemark{i}}
}
\startdata
  33 & 16:06:15.69 & 55:28:01.7 & 1276.2 &  171.7 & $0.48_{-0.20}^{+0.20}$ & $22.46\pm  0.01$ & $21.77\pm  0.01$ & $21.32\pm  0.00$ &   0.49 & 0.04 \\
 102 & 16:06:25.37 & 55:26:53.4 & 3415.4 &  234.9 & $1.22_{-0.29}^{+0.29}$ & $25.01\pm  0.07$ & $23.57\pm  0.02$ & $21.96\pm  0.01$ &   0.29 & 0.03 \\
 127 & 16:06:28.55 & 55:26:30.0 & 4128.2 &  270.6 & $0.66_{-0.22}^{+0.22}$ & $22.98\pm  0.01$ & $22.42\pm  0.01$ & $21.77\pm  0.00$ &   1.13 & 0.03 \\
 198 & 16:06:23.32 & 55:27:01.5 & 3039.8 &  322.6 & $1.35_{-0.31}^{+0.31}$ & $25.32\pm  0.03$ & $25.44\pm  0.03$ & $25.42\pm  0.03$ &   0.14 & 0.34 \\
 201 & 16:06:21.53 & 55:27:13.1 & 2657.3 &  326.1 & $1.86_{-0.38}^{+0.38}$ & $25.22\pm  0.02$ & $25.45\pm  0.03$ & $25.66\pm  0.03$ &   0.15 & 0.15 \\
 241 & 16:06:18.27 & 55:27:22.3 & 2107.1 &  524.0 & $0.38_{-0.18}^{+0.18}$ & $21.61\pm  0.01$ & $20.73\pm  0.00$ & $20.28\pm  0.00$ &   0.25 & 0.03 \\
 275 & 16:06:19.79 & 55:27:18.6 & 2356.3 &  422.5 & $0.60_{-0.21}^{+0.21}$ & $25.70\pm  0.06$ & $25.16\pm  0.03$ & $24.61\pm  0.02$ &   1.37 & 0.00 \\
 297 & 16:06:15.65 & 55:27:44.9 & 1477.8 &  441.5 & $0.72_{-0.22}^{+0.22}$ & $26.80\pm  0.11$ & $26.03\pm  0.06$ & $25.09\pm  0.03$ &   0.70 & 0.00 \\
 343 & 16:06:23.45 & 55:26:49.0 & 3211.7 &  507.1 & $0.52_{-0.20}^{+0.20}$ & $23.36\pm  0.01$ & $22.63\pm  0.01$ & $22.11\pm  0.01$ &   0.14 & 0.04 \\
 347 & 16:06:14.94 & 55:27:46.1 & 1367.3 &  496.4 & $1.38_{-0.31}^{+0.31}$ & $23.57\pm  0.01$ & $23.74\pm  0.01$ & $23.77\pm  0.01$ &   0.75 & 0.03 \\
 389 & 16:06:21.45 & 55:26:57.2 & 2842.8 &  585.7 & $0.69_{-0.22}^{+0.22}$ & $23.55\pm  0.02$ & $22.55\pm  0.01$ & $21.48\pm  0.00$ &   0.40 & 0.03 \\
 416 & 16:06:12.27 & 55:27:58.3 &  860.0 &  584.5 & $0.80_{-0.24}^{+0.24}$ & $25.20\pm  0.04$ & $24.71\pm  0.03$ & $23.80\pm  0.01$ &   1.07 & 0.00 \\
 442 & 16:06:21.32 & 55:26:56.1 & 2838.4 &  617.9 & $1.87_{-0.38}^{+0.38}$ & $25.63\pm  0.03$ & $26.01\pm  0.04$ & $26.31\pm  0.06$ &   0.34 & 0.03 \\
 446 & 16:06:01.09 & 55:24:57.7 & 1579.7 & 4606.0 & $0.67_{-0.22}^{+0.22}$ & $25.68\pm  0.06$ & $24.55\pm  0.02$ & $23.35\pm  0.01$ &   0.27 & 0.03 \\
 467 & 16:06:01.41 & 55:24:55.9 & 1644.9 & 4602.1 & $0.86_{-0.24}^{+0.24}$ & $24.62\pm  0.04$ & $23.50\pm  0.01$ & $22.18\pm  0.01$ &   0.22 & 0.03 \\
 471 & 16:06:00.96 & 55:24:58.8 & 1548.5 & 4603.4 & $2.75_{-0.49}^{+0.49}$ & $24.05\pm  0.02$ & $23.96\pm  0.01$ & $24.28\pm  0.02$ &   0.76 & 0.03 \\
 584 & 16:06:02.42 & 55:24:49.5 & 1859.5 & 4596.8 & $0.51_{-0.20}^{+0.20}$ & $23.19\pm  0.01$ & $22.12\pm  0.01$ & $21.33\pm  0.00$ &   0.26 & 0.03 \\
 677 & 16:06:03.96 & 55:24:52.2 & 2032.8 & 4393.1 & $0.76_{-0.23}^{+0.23}$ & $26.21\pm  0.08$ & $25.39\pm  0.04$ & $24.32\pm  0.02$ &   0.23 & 0.03 \\
 689 & 16:06:03.68 & 55:24:56.0 & 1948.9 & 4363.0 & $0.98_{-0.26}^{+0.26}$ & $24.03\pm  0.03$ & $23.15\pm  0.01$ & $21.87\pm  0.01$ &   0.74 & 0.02 \\
 725 & 16:06:14.40 & 55:23:46.6 & 4241.3 & 4333.3 & $0.68_{-0.22}^{+0.22}$ & $26.58\pm  0.06$ & $25.89\pm  0.03$ & $25.12\pm  0.02$ &   0.24 & 0.03 \\
 737 & 16:06:03.49 & 55:24:57.6 & 1903.3 & 4355.7 & $0.64_{-0.22}^{+0.21}$ & $25.19\pm  0.04$ & $24.63\pm  0.03$ & $24.01\pm  0.02$ &   0.51 & 0.03 \\
 796 & 16:05:58.07 & 55:25:38.2 &  676.5 & 4282.4 & $1.87_{-0.38}^{+0.38}$ & $23.74\pm  0.02$ & $23.99\pm  0.02$ & $24.31\pm  0.03$ &   3.53 & 0.00 \\
 813 & 16:06:14.11 & 55:23:48.9 & 4173.7 & 4328.4 & $0.33_{-0.17}^{+0.17}$ & $22.72\pm  0.01$ & $22.19\pm  0.00$ & $21.96\pm  0.00$ &   1.76 & 0.03 \\
 819 & 16:06:04.99 & 55:24:42.7 & 2288.0 & 4434.4 & $0.76_{-0.23}^{+0.23}$ & $24.28\pm  0.03$ & $23.30\pm  0.01$ & $22.06\pm  0.01$ &   0.55 & 0.03 \\
 892 & 16:06:10.86 & 55:24:16.8 & 3394.6 & 4227.8 & $1.11_{-0.28}^{+0.28}$ & $24.85\pm  0.02$ & $24.95\pm  0.02$ & $24.66\pm  0.02$ &   0.47 & 0.03 \\
 960 & 16:06:13.49 & 55:24:03.5 & 3911.4 & 4163.1 & $0.58_{-0.21}^{+0.21}$ & $24.92\pm  0.03$ & $23.81\pm  0.01$ & $22.76\pm  0.01$ &   0.91 & 0.03 \\
1135 & 16:06:16.61 & 55:23:50.7 & 4488.1 & 4035.9 & $1.86_{-0.38}^{+0.38}$ & $24.44\pm  0.02$ & $24.67\pm  0.02$ & $24.90\pm  0.02$ &   0.51 & 0.03 \\
1266 & 16:05:58.73 & 55:25:58.5 &  515.4 & 3893.4 & $1.25_{-0.30}^{+0.29}$ & $25.45\pm  0.03$ & $25.61\pm  0.03$ & $25.43\pm  0.03$ &   0.16 & 0.03 \\
1342 & 16:06:12.46 & 55:24:33.3 & 3406.5 & 3800.0 & $2.74_{-0.49}^{+0.49}$ & $25.78\pm  0.02$ & $25.67\pm  0.02$ & $26.04\pm  0.03$ &   0.17 & 0.07 \\
1371 & 16:06:14.90 & 55:24:16.9 & 3936.3 & 3802.8 & $1.26_{-0.30}^{+0.30}$ & $24.41\pm  0.01$ & $24.52\pm  0.01$ & $24.39\pm  0.01$ &   0.59 & 0.03 \\
1455 & 16:06:13.24 & 55:24:34.1 & 3500.8 & 3705.5 & $0.58_{-0.21}^{+0.21}$ & $25.88\pm  0.03$ & $25.17\pm  0.02$ & $24.57\pm  0.01$ &   0.17 & 0.04 \\
1517 & 16:05:59.95 & 55:26:06.7 &  578.9 & 3637.0 & $0.71_{-0.22}^{+0.23}$ & $24.18\pm  0.03$ & $23.47\pm  0.01$ & $22.60\pm  0.01$ &   2.60 & 0.03 \\
1520 & 16:06:00.04 & 55:26:06.0 &  599.6 & 3638.3 & $0.62_{-0.21}^{+0.21}$ & $26.04\pm  0.07$ & $25.20\pm  0.03$ & $24.42\pm  0.02$ &   0.81 & 0.00 \\
1532 & 16:06:01.17 & 55:26:01.0 &  812.3 & 3598.8 & $0.67_{-0.22}^{+0.22}$ & $22.76\pm  0.01$ & $22.14\pm  0.01$ & $21.42\pm  0.00$ &   0.40 & 0.03 \\
1538 & 16:06:16.15 & 55:24:19.8 & 4068.7 & 3625.5 & $1.86_{-0.38}^{+0.38}$ & $24.50\pm  0.01$ & $24.70\pm  0.01$ & $24.95\pm  0.02$ &   0.15 & 0.34 \\
1558 & 16:06:00.33 & 55:26:07.5 &  620.4 & 3582.9 & $0.74_{-0.23}^{+0.23}$ & $24.43\pm  0.03$ & $23.57\pm  0.02$ & $22.50\pm  0.01$ &   0.79 & 0.03 \\
1572 & 16:06:11.07 & 55:24:56.5 & 2935.9 & 3579.6 & $0.52_{-0.20}^{+0.20}$ & $25.64\pm  0.04$ & $25.08\pm  0.02$ & $24.66\pm  0.02$ &   0.28 & 0.02 \\
1580 & 16:06:15.61 & 55:24:27.1 & 3905.4 & 3567.3 & $0.62_{-0.21}^{+0.21}$ & $24.04\pm  0.02$ & $23.54\pm  0.01$ & $22.98\pm  0.01$ &   2.05 & 0.00 \\
1718 & 16:05:59.80 & 55:26:14.2 &  466.2 & 3533.5 & $0.85_{-0.24}^{+0.24}$ & $23.86\pm  0.03$ & $22.60\pm  0.01$ & $21.19\pm  0.00$ &   0.19 & 0.03 \\
1746 & 16:06:15.50 & 55:24:37.5 & 3763.7 & 3415.4 & $0.63_{-0.22}^{+0.21}$ & $23.15\pm  0.01$ & $22.77\pm  0.01$ & $22.35\pm  0.01$ &   2.40 & 0.00 \\
1827 & 16:06:11.99 & 55:25:04.1 & 2965.7 & 3363.4 & $1.39_{-0.31}^{+0.31}$ & $25.36\pm  0.02$ & $25.56\pm  0.02$ & $25.57\pm  0.03$ &   0.37 & 0.03 \\
1921 & 16:06:03.72 & 55:26:05.6 & 1097.7 & 3258.4 & $1.04_{-0.27}^{+0.27}$ & $24.28\pm  0.02$ & $23.32\pm  0.01$ & $21.97\pm  0.01$ &   0.10 & 0.92 \\
2079 & 16:06:12.58 & 55:25:12.9 & 2936.0 & 3161.4 & $1.19_{-0.29}^{+0.29}$ & $25.09\pm  0.02$ & $25.24\pm  0.02$ & $25.01\pm  0.02$ &   0.25 & 0.03 \\
2155 & 16:06:02.66 & 55:26:17.8 &  806.4 & 3176.8 & $0.88_{-0.25}^{+0.25}$ & $24.24\pm  0.02$ & $23.88\pm  0.01$ & $22.98\pm  0.01$ &   0.40 & 0.03 \\
2156 & 16:06:03.16 & 55:26:16.4 &  889.8 & 3146.4 & $1.33_{-0.31}^{+0.31}$ & $23.64\pm  0.01$ & $23.77\pm  0.01$ & $23.79\pm  0.01$ &   0.83 & 0.03 \\
2499 & 16:06:15.18 & 55:25:13.6 & 3276.3 & 2878.5 & $0.52_{-0.20}^{+0.20}$ & $21.75\pm  0.01$ & $20.96\pm  0.00$ & $20.43\pm  0.00$ &   5.03 & 0.03 \\
2685 & 16:06:15.22 & 55:25:16.7 & 3243.3 & 2826.5 & $1.38_{-0.31}^{+0.31}$ & $24.58\pm  0.02$ & $24.71\pm  0.02$ & $24.70\pm  0.02$ &   1.12 & 0.00 \\
2694 & 16:06:10.76 & 55:25:49.4 & 2243.1 & 2776.5 & $0.07_{-0.07}^{+0.14}$ & $20.60\pm  0.00$ & $20.54\pm  0.00$ & $20.68\pm  0.00$ &   4.30 & 0.03 \\
2696 & 16:06:19.89 & 55:24:39.7 & 4325.1 & 2920.1 & $1.24_{-0.29}^{+0.29}$ & $25.38\pm  0.03$ & $25.51\pm  0.04$ & $25.35\pm  0.03$ &   0.94 & 0.03 \\
2730 & 16:06:06.50 & 55:26:17.5 & 1324.6 & 2779.8 & $1.29_{-0.30}^{+0.30}$ & $24.91\pm  0.02$ & $24.98\pm  0.02$ & $24.94\pm  0.02$ &   0.75 & 0.03 \\
2736 & 16:06:05.25 & 55:26:25.2 & 1063.2 & 2789.5 & $0.70_{-0.22}^{+0.22}$ & $23.96\pm  0.02$ & $23.15\pm  0.01$ & $22.26\pm  0.00$ &   0.49 & 0.03 \\
2785 & 16:06:05.67 & 55:26:26.0 & 1109.1 & 2732.7 & $0.35_{-0.18}^{+0.18}$ & $22.35\pm  0.01$ & $21.78\pm  0.00$ & $21.55\pm  0.00$ &   1.73 & 0.03 \\
2845 & 16:06:11.81 & 55:25:47.1 & 2412.1 & 2702.8 & $0.69_{-0.22}^{+0.22}$ & $23.05\pm  0.01$ & $22.74\pm  0.01$ & $22.30\pm  0.00$ &   0.69 & 0.03 \\
3002 & 16:06:03.76 & 55:26:47.4 &  591.2 & 2594.5 & $1.87_{-0.38}^{+0.38}$ & $25.06\pm  0.03$ & $25.31\pm  0.03$ & $25.76\pm  0.05$ &   1.06 & 0.03 \\
3017 & 16:06:04.55 & 55:26:43.8 &  740.8 & 2568.6 & $0.26_{-0.17}^{+0.17}$ & $22.23\pm  0.00$ & $21.95\pm  0.00$ & $21.97\pm  0.00$ &   0.30 & 0.03 \\
3091 & 16:06:08.44 & 55:26:20.7 & 1546.0 & 2526.0 & $1.31_{-0.30}^{+0.30}$ & $26.54\pm  0.10$ & $25.45\pm  0.04$ & $23.75\pm  0.01$ &   0.26 & 0.03 \\
3094 & 16:06:10.55 & 55:26:09.4 & 1967.6 & 2483.0 & $0.52_{-0.20}^{+0.20}$ & $22.07\pm  0.01$ & $21.15\pm  0.00$ & $20.55\pm  0.00$ &   3.97 & 0.03 \\
3108 & 16:06:07.92 & 55:26:26.2 & 1408.9 & 2494.6 & $0.60_{-0.21}^{+0.21}$ & $23.79\pm  0.02$ & $23.27\pm  0.01$ & $22.75\pm  0.01$ &   1.93 & 0.03 \\
3117 & 16:06:03.03 & 55:26:58.1 &  359.8 & 2501.5 & $0.90_{-0.25}^{+0.25}$ & $23.70\pm  0.02$ & $23.28\pm  0.01$ & $22.26\pm  0.01$ &   0.13 & 0.04 \\
3380 & 16:06:10.46 & 55:26:25.6 & 1757.2 & 2236.7 & $0.29_{-0.17}^{+0.17}$ & $21.78\pm  0.00$ & $21.05\pm  0.00$ & $20.66\pm  0.00$ &   2.04 & 0.03 \\
3392 & 16:06:21.09 & 55:25:14.4 & 4059.9 & 2245.5 & $1.26_{-0.30}^{+0.30}$ & $25.24\pm  0.03$ & $25.36\pm  0.03$ & $25.23\pm  0.03$ &   2.08 & 0.00 \\
3414 & 16:06:19.54 & 55:25:26.0 & 3709.0 & 2224.7 & $0.57_{-0.21}^{+0.21}$ & $23.41\pm  0.01$ & $22.92\pm  0.01$ & $22.47\pm  0.01$ &   0.89 & 0.03 \\
3429 & 16:06:12.02 & 55:26:16.3 & 2081.6 & 2220.1 & $1.30_{-0.30}^{+0.30}$ & $23.89\pm  0.01$ & $23.97\pm  0.01$ & $23.92\pm  0.01$ &   2.09 & 0.00 \\
3480 & 16:06:08.01 & 55:26:45.4 & 1184.9 & 2180.0 & $1.88_{-0.38}^{+0.38}$ & $24.39\pm  0.01$ & $24.63\pm  0.01$ & $25.12\pm  0.02$ &   0.65 & 0.03 \\
3486 & 16:06:14.13 & 55:26:05.5 & 2496.7 & 2170.3 & $0.77_{-0.23}^{+0.23}$ & $24.91\pm  0.02$ & $24.44\pm  0.01$ & $23.62\pm  0.01$ &   0.36 & 0.03 \\
3553 & 16:06:08.05 & 55:26:52.6 & 1100.9 & 2062.5 & $0.26_{-0.17}^{+0.16}$ & $21.67\pm  0.00$ & $21.37\pm  0.00$ & $21.32\pm  0.00$ &   5.35 & 0.00 \\
3588 & 16:06:11.48 & 55:26:30.4 & 1835.5 & 2053.9 & $0.12_{-0.12}^{+0.15}$ & $22.84\pm  0.01$ & $22.71\pm  0.01$ & $22.71\pm  0.01$ &   0.26 & 0.03 \\
3597 & 16:06:21.03 & 55:25:26.4 & 3904.3 & 2062.6 & $1.36_{-0.31}^{+0.31}$ & $25.73\pm  0.03$ & $25.93\pm  0.03$ & $25.91\pm  0.03$ &   0.24 & 0.02 \\
3654 & 16:06:15.58 & 55:26:07.3 & 2669.8 & 1989.5 & $0.62_{-0.21}^{+0.21}$ & $23.22\pm  0.01$ & $22.79\pm  0.01$ & $22.31\pm  0.01$ &   2.60 & 0.00 \\
3915 & 16:06:17.02 & 55:26:13.2 & 2790.1 & 1746.0 & $1.15_{-0.28}^{+0.28}$ & $25.51\pm  0.02$ & $25.62\pm  0.02$ & $25.37\pm  0.02$ &   0.42 & 0.03 \\
3918 & 16:06:24.61 & 55:25:23.1 & 4424.8 & 1739.7 & $1.24_{-0.29}^{+0.29}$ & $24.54\pm  0.02$ & $24.63\pm  0.02$ & $24.49\pm  0.02$ &   0.53 & 0.03 \\
3982 & 16:06:22.56 & 55:25:30.3 & 4060.7 & 1840.8 & $1.88_{-0.38}^{+0.38}$ & $25.13\pm  0.02$ & $25.40\pm  0.02$ & $25.81\pm  0.03$ &   0.19 & 0.03 \\
3985 & 16:06:12.17 & 55:26:50.1 & 1685.4 & 1671.9 & $0.66_{-0.22}^{+0.22}$ & $24.39\pm  0.02$ & $23.92\pm  0.01$ & $23.31\pm  0.01$ &   1.51 & 0.03 \\
4041 & 16:06:13.37 & 55:26:45.1 & 1907.3 & 1623.3 & $1.24_{-0.29}^{+0.29}$ & $25.73\pm  0.03$ & $25.92\pm  0.04$ & $25.71\pm  0.03$ &   0.79 & 0.02 \\
4056 & 16:06:24.69 & 55:25:33.2 & 4310.5 & 1572.4 & $0.60_{-0.21}^{+0.21}$ & $23.94\pm  0.03$ & $23.08\pm  0.01$ & $22.34\pm  0.01$ &   2.74 & 0.00 \\
4084 & 16:06:16.86 & 55:26:23.5 & 2642.4 & 1599.4 & $1.35_{-0.31}^{+0.31}$ & $25.76\pm  0.03$ & $25.94\pm  0.03$ & $25.89\pm  0.03$ &   0.24 & 0.03 \\
4172 & 16:06:10.41 & 55:27:10.0 & 1204.6 & 1539.9 & $0.52_{-0.20}^{+0.20}$ & $26.16\pm  0.06$ & $24.99\pm  0.02$ & $24.07\pm  0.01$ &   0.82 & 0.00 \\
4176 & 16:06:07.71 & 55:27:10.0 &  843.4 & 1824.2 & $0.73_{-0.23}^{+0.23}$ & $25.15\pm  0.02$ & $24.75\pm  0.02$ & $24.09\pm  0.01$ &   0.40 & 0.03 \\
4213 & 16:06:17.45 & 55:26:26.2 & 2687.3 & 1495.1 & $1.18_{-0.29}^{+0.29}$ & $25.58\pm  0.03$ & $25.72\pm  0.03$ & $25.48\pm  0.03$ &   1.14 & 0.00 \\
4227 & 16:06:17.06 & 55:26:29.8 & 2590.5 & 1478.1 & $0.80_{-0.24}^{+0.24}$ & $23.26\pm  0.01$ & $22.93\pm  0.01$ & $22.21\pm  0.01$ &   1.46 & 0.03 \\
4329 & 16:06:15.08 & 55:26:48.7 & 2092.5 & 1387.8 & $0.39_{-0.18}^{+0.18}$ & $24.22\pm  0.01$ & $23.92\pm  0.01$ & $24.16\pm  0.01$ &   1.52 & 0.03 \\
4404 & 16:06:15.33 & 55:26:50.7 & 2101.8 & 1329.8 & $0.52_{-0.20}^{+0.20}$ & $22.82\pm  0.01$ & $22.11\pm  0.00$ & $21.60\pm  0.00$ &   0.28 & 0.03 \\
4447 & 16:06:26.56 & 55:25:37.5 & 4509.2 & 1308.5 & $1.87_{-0.38}^{+0.38}$ & $25.43\pm  0.03$ & $25.81\pm  0.04$ & $26.14\pm  0.06$ &   0.74 & 0.07 \\
4489 & 16:06:11.78 & 55:27:20.1 & 1263.1 & 1237.7 & $0.27_{-0.17}^{+0.17}$ & $22.14\pm  0.01$ & $21.83\pm  0.00$ & $21.78\pm  0.00$ &   0.67 & 0.03 \\
4572 & 16:06:13.50 & 55:27:12.0 & 1594.9 & 1185.2 & $1.86_{-0.38}^{+0.38}$ & $25.05\pm  0.02$ & $25.24\pm  0.02$ & $25.52\pm  0.03$ &   0.70 & 0.03 \\
4684 & 16:06:14.02 & 55:27:15.8 & 1617.1 & 1071.3 & $0.62_{-0.21}^{+0.21}$ & $23.45\pm  0.01$ & $23.03\pm  0.01$ & $22.55\pm  0.01$ &   0.47 & 0.03 \\
4777 & 16:06:09.91 & 55:27:46.7 &  686.5 & 1013.4 & $1.15_{-0.28}^{+0.28}$ & $24.43\pm  0.01$ & $24.54\pm  0.02$ & $24.29\pm  0.01$ &   0.64 & 0.03 \\
4789 & 16:06:15.33 & 55:27:10.6 & 1857.7 & 1015.7 & $0.77_{-0.23}^{+0.23}$ & $25.62\pm  0.04$ & $24.25\pm  0.01$ & $22.82\pm  0.01$ &   0.19 & 0.03 \\
4813 & 16:06:09.70 & 55:27:52.0 &  593.7 &  952.5 & $0.36_{-0.18}^{+0.18}$ & $22.06\pm  0.01$ & $21.25\pm  0.00$ & $20.83\pm  0.00$ &   0.18 & 0.03 \\
4832 & 16:06:10.98 & 55:27:44.1 &  861.4 &  942.8 & $0.50_{-0.20}^{+0.20}$ & $22.58\pm  0.01$ & $21.73\pm  0.00$ & $21.18\pm  0.00$ &   0.24 & 0.03 \\
4877 & 16:06:27.89 & 55:25:52.7 & 4499.9 &  928.4 & $0.69_{-0.22}^{+0.22}$ & $25.07\pm  0.04$ & $24.64\pm  0.03$ & $24.03\pm  0.02$ &   1.37 & 0.02 \\
4941 & 16:06:11.72 & 55:27:43.3 &  970.8 &  878.6 & $0.81_{-0.24}^{+0.24}$ & $24.36\pm  0.03$ & $22.97\pm  0.01$ & $21.46\pm  0.00$ &   0.20 & 0.03 \\
4960 & 16:06:28.09 & 55:25:55.5 & 4492.1 &  863.6 & $0.39_{-0.18}^{+0.18}$ & $23.38\pm  0.01$ & $22.99\pm  0.01$ & $22.97\pm  0.01$ &   2.27 & 0.03 \\
4966 & 16:06:26.99 & 55:26:02.4 & 4259.2 &  869.5 & $1.86_{-0.38}^{+0.38}$ & $24.88\pm  0.02$ & $25.12\pm  0.02$ & $25.38\pm  0.03$ &   0.30 & 0.03 \\
5077 & 16:06:11.57 & 55:27:50.1 &  866.5 &  785.6 & $1.88_{-0.38}^{+0.38}$ & $25.06\pm  0.02$ & $25.31\pm  0.02$ & $25.74\pm  0.03$ &   0.22 & 0.03 \\
5097 & 16:06:13.42 & 55:27:38.3 & 1261.2 &  779.4 & $1.12_{-0.28}^{+0.28}$ & $25.34\pm  0.02$ & $25.45\pm  0.02$ & $25.17\pm  0.02$ &   0.20 & 0.03 \\
5127 & 16:06:17.31 & 55:27:14.8 & 2070.5 &  742.0 & $0.62_{-0.21}^{+0.21}$ & $25.40\pm  0.04$ & $24.85\pm  0.02$ & $24.26\pm  0.02$ &   1.63 & 0.00 \\
5173 & 16:06:20.96 & 55:26:52.2 & 2838.9 &  716.1 & $1.87_{-0.38}^{+0.38}$ & $24.95\pm  0.02$ & $25.20\pm  0.03$ & $25.60\pm  0.04$ &   1.10 & 0.03 \\
5182 & 16:06:11.37 & 55:27:57.4 &  750.8 &  692.1 & $0.75_{-0.23}^{+0.23}$ & $25.56\pm  0.05$ & $24.98\pm  0.03$ & $24.11\pm  0.02$ &   0.96 & 0.00 \\
5239 & 16:06:13.34 & 55:27:46.9 & 1144.2 &  652.2 & $1.33_{-0.31}^{+0.31}$ & $25.08\pm  0.03$ & $25.30\pm  0.03$ & $25.21\pm  0.03$ &   1.28 & 0.03 \\
5256 & 16:06:15.42 & 55:27:33.5 & 1587.5 &  644.3 & $4.46_{-0.72}^{+0.72}$ & $29.01\pm  0.42$ & $26.74\pm  0.06$ & $25.86\pm  0.04$ &   0.26 & 0.02 \\
5355 & 16:06:04.68 & 55:24:18.2 & 2548.7 & 4854.3 & $0.72_{-0.22}^{+0.22}$ & $26.36\pm  0.08$ & $25.63\pm  0.04$ & $24.73\pm  0.02$ &   0.33 & 0.03 \\
5617 & 16:05:58.02 & 55:25:10.8 & 1006.8 & 4720.5 & $0.61_{-0.21}^{+0.21}$ & $25.76\pm  0.04$ & $25.08\pm  0.02$ & $24.43\pm  0.02$ &   0.24 & 0.53 
\enddata
\tablenotetext{a}{SExtractor ID}
\tablenotetext{b}{Right Ascension (J2000)}
\tablenotetext{c}{Declination (J2000)}
\tablenotetext{d}{Bayesian photometric redshift}
\tablenotetext{e}{AB Isophotal magnitude in the F475W filter}
\tablenotetext{f}{AB Isophotal magnitude in the F606W filter}
\tablenotetext{g}{AB Isophotal magnitude in the F814W filter}
\tablenotetext{h}{Full width at half maximum as measured by $SExtractor$ in arcsec}
\tablenotetext{i}{SExtractor star/galaxy classification}
\tablecomments{Catalog with magnitudes and photometric redshifts in the VV~29 field. Only
galaxies outside the masked area with $I_{F814W} < 26$, and very high values of the Bayesian odds ($>0.99$) are
included. This is the subsample of galaxies for which the photometric redshifts are most reliable.
The full catalog published electronically contains more information about these objects and about
the rest of the detections in the field.}
\end{deluxetable}

\clearpage

\begin{deluxetable}{lrrrrrrrrrr}
\label{table_NGC 4676} 
\tabletypesize{\scriptsize}
\tablecaption{Photometric Redshift catalog in the NGC 4676 field}
\tablewidth{0pt} 
\tablehead{
\colhead{ID \tablenotemark{a}}&
\colhead{Ra \tablenotemark{b}}&
\colhead{Dec \tablenotemark{c}}&
\colhead{X}&
\colhead{Y}&
\colhead{$z_b$ \tablenotemark{d}}&
\colhead{$g_{F475W}$ \tablenotemark{e}}&
\colhead{$V_{F606W}$ \tablenotemark{f}}&
\colhead{$I_{F814W}$ \tablenotemark{g}}&
\colhead{FWHM \tablenotemark{h}}&
\colhead{s \tablenotemark{i}}
}
\startdata
 125 & 12:46:16.88 & 30:43:31.2 & 3425.2 & 3174.8 & $3.66_{-0.61}^{+0.61}$ & $26.56\pm  0.06$ & $25.79\pm  0.03$ & $25.71\pm  0.03$ &   0.13 & 0.45 \\
 167 & 12:46:14.50 & 30:42:23.2 & 2259.7 & 4108.7 & $1.13_{-0.28}^{+0.28}$ & $24.07\pm  0.02$ & $24.19\pm  0.02$ & $23.90\pm  0.01$ &   0.61 & 0.03 \\
 182 & 12:46:17.44 & 30:42:32.3 & 3018.0 & 4290.3 & $0.53_{-0.20}^{+0.20}$ & $24.78\pm  0.03$ & $23.65\pm  0.01$ & $22.73\pm  0.01$ &   0.18 & 0.03 \\
 205 & 12:46:17.91 & 30:43:42.8 & 3765.8 & 3089.1 & $0.50_{-0.20}^{+0.20}$ & $23.28\pm  0.01$ & $22.08\pm  0.01$ & $21.20\pm  0.01$ &   0.20 & 0.03 \\
 236 & 12:46:19.76 & 30:43:26.1 & 4038.6 & 3602.5 & $0.54_{-0.20}^{+0.20}$ & $24.41\pm  0.02$ & $23.63\pm  0.01$ & $23.06\pm  0.01$ &   0.26 & 0.03 \\
 238 & 12:46:17.25 & 30:43:38.3 & 3572.9 & 3092.1 & $1.29_{-0.30}^{+0.30}$ & $24.98\pm  0.03$ & $25.17\pm  0.03$ & $25.01\pm  0.03$ &   0.81 & 0.03 \\
 278 & 12:46:16.98 & 30:43:41.9 & 3544.5 & 2996.1 & $0.59_{-0.21}^{+0.21}$ & $22.25\pm  0.01$ & $21.32\pm  0.00$ & $20.57\pm  0.00$ &   1.25 & 0.03 \\
 335 & 12:46:20.10 & 30:43:16.8 & 4033.2 & 3808.1 & $1.11_{-0.28}^{+0.28}$ & $25.47\pm  0.04$ & $25.65\pm  0.04$ & $25.29\pm  0.03$ &   0.61 & 0.00 \\
 461 & 12:46:21.27 & 30:43:11.9 & 4258.2 & 4031.3 & $0.79_{-0.23}^{+0.23}$ & $23.22\pm  0.01$ & $22.65\pm  0.01$ & $21.70\pm  0.01$ &   0.31 & 0.03 \\
 543 & 12:46:18.84 & 30:44:05.1 & 4182.3 & 2799.8 & $0.27_{-0.17}^{+0.17}$ & $22.72\pm  0.01$ & $22.43\pm  0.01$ & $22.47\pm  0.01$ &   3.77 & 0.00 \\
 576 & 12:46:05.58 & 30:42:42.1 &  381.8 & 2726.5 & $0.34_{-0.18}^{+0.18}$ & $23.06\pm  0.01$ & $22.68\pm  0.01$ & $22.64\pm  0.01$ &   0.66 & 0.03 \\
 577 & 12:46:08.45 & 30:42:02.7 &  682.9 & 3765.5 & $0.64_{-0.22}^{+0.21}$ & $26.72\pm  0.11$ & $25.98\pm  0.05$ & $25.24\pm  0.03$ &   0.40 & 0.00 \\
 578 & 12:46:13.76 & 30:43:30.3 & 2699.7 & 2826.6 & $0.75_{-0.23}^{+0.23}$ & $26.47\pm  0.09$ & $25.63\pm  0.04$ & $24.56\pm  0.02$ &   0.15 & 0.03 \\
 614 & 12:46:17.82 & 30:44:02.9 & 3928.8 & 2720.1 & $1.17_{-0.28}^{+0.29}$ & $25.57\pm  0.03$ & $25.70\pm  0.03$ & $25.45\pm  0.03$ &   0.17 & 0.13 \\
 625 & 12:46:20.56 & 30:42:54.6 & 3938.7 & 4256.9 & $0.75_{-0.23}^{+0.23}$ & $25.20\pm  0.04$ & $24.65\pm  0.03$ & $23.79\pm  0.01$ &   1.58 & 0.00 \\
 719 & 12:46:13.23 & 30:43:40.8 & 2671.3 & 2576.7 & $0.33_{-0.18}^{+0.17}$ & $22.69\pm  0.01$ & $22.32\pm  0.01$ & $22.27\pm  0.01$ &   1.45 & 0.00 \\
 769 & 12:46:17.02 & 30:44:08.3 & 3793.6 & 2529.5 & $1.41_{-0.32}^{+0.32}$ & $24.51\pm  0.02$ & $24.71\pm  0.02$ & $24.74\pm  0.02$ &   0.37 & 0.05 \\
 814 & 12:46:06.03 & 30:42:58.1 &  631.8 & 2494.0 & $1.94_{-0.39}^{+0.39}$ & $24.71\pm  0.02$ & $24.96\pm  0.02$ & $25.16\pm  0.03$ &   0.18 & 0.03 \\
 825 & 12:46:14.84 & 30:42:32.0 & 2417.3 & 3990.5 & $1.88_{-0.38}^{+0.38}$ & $25.13\pm  0.02$ & $25.48\pm  0.03$ & $25.98\pm  0.05$ &   0.59 & 0.02 \\
 833 & 12:46:17.67 & 30:43:19.0 & 3495.3 & 3485.6 & $0.49_{-0.20}^{+0.20}$ & $22.44\pm  0.01$ & $21.78\pm  0.00$ & $21.34\pm  0.00$ &   0.44 & 0.03 \\
 838 & 12:46:15.32 & 30:44:01.5 & 3340.1 & 2452.3 & $0.49_{-0.20}^{+0.20}$ & $23.51\pm  0.02$ & $22.37\pm  0.01$ & $21.53\pm  0.00$ &   0.30 & 0.03 \\
 872 & 12:46:13.00 & 30:43:47.5 & 2680.8 & 2429.9 & $1.30_{-0.30}^{+0.30}$ & $24.64\pm  0.03$ & $24.81\pm  0.03$ & $24.68\pm  0.03$ &   1.32 & 0.00 \\
 982 & 12:46:05.70 & 30:43:04.2 &  609.7 & 2347.2 & $3.52_{-0.59}^{+0.59}$ & $25.72\pm  0.04$ & $25.06\pm  0.02$ & $25.08\pm  0.03$ &   0.36 & 0.03 \\
1176 & 12:46:08.35 & 30:41:55.2 &  593.7 & 3886.6 & $0.62_{-0.21}^{+0.21}$ & $20.78\pm  0.00$ & $20.36\pm  0.00$ & $19.90\pm  0.00$ &   0.28 & 0.03 \\
1201 & 12:46:13.38 & 30:44:09.7 & 2970.9 & 2079.1 & $0.54_{-0.20}^{+0.20}$ & $22.79\pm  0.01$ & $22.04\pm  0.01$ & $21.49\pm  0.00$ &   1.37 & 0.03 \\
1289 & 12:46:04.04 & 30:43:14.3 &  321.9 & 1973.3 & $0.54_{-0.20}^{+0.20}$ & $21.22\pm  0.00$ & $20.61\pm  0.00$ & $20.11\pm  0.00$ &   0.82 & 0.03 \\
1317 & 12:46:14.44 & 30:44:21.8 & 3322.1 & 1987.7 & $0.49_{-0.20}^{+0.20}$ & $24.26\pm  0.03$ & $23.34\pm  0.01$ & $22.76\pm  0.01$ &   1.82 & 0.00 \\
1346 & 12:46:18.39 & 30:42:45.7 & 3357.6 & 4162.0 & $1.88_{-0.38}^{+0.38}$ & $24.72\pm  0.02$ & $24.96\pm  0.02$ & $25.37\pm  0.03$ &   0.14 & 0.72 \\
1356 & 12:46:13.15 & 30:44:14.8 & 2962.0 & 1961.7 & $0.68_{-0.22}^{+0.22}$ & $25.81\pm  0.06$ & $25.25\pm  0.04$ & $24.55\pm  0.02$ &   0.90 & 0.00 \\
1411 & 12:46:18.46 & 30:43:29.6 & 3774.1 & 3388.3 & $0.53_{-0.20}^{+0.20}$ & $23.08\pm  0.01$ & $22.23\pm  0.01$ & $21.63\pm  0.01$ &   0.26 & 0.03 \\
1459 & 12:46:18.22 & 30:42:57.6 & 3427.2 & 3929.2 & $0.58_{-0.21}^{+0.21}$ & $23.16\pm  0.01$ & $22.74\pm  0.01$ & $22.38\pm  0.01$ &   0.18 & 0.03 \\
1518 & 12:46:20.68 & 30:42:53.4 & 3954.6 & 4293.1 & $0.41_{-0.18}^{+0.18}$ & $23.75\pm  0.02$ & $22.51\pm  0.01$ & $21.81\pm  0.01$ &   0.18 & 0.03 \\
1528 & 12:46:10.17 & 30:42:18.8 & 1226.4 & 3679.0 & $0.58_{-0.21}^{+0.21}$ & $24.57\pm  0.04$ & $23.55\pm  0.02$ & $22.63\pm  0.01$ &   0.73 & 0.01 \\
1561 & 12:46:08.71 & 30:41:56.8 &  690.8 & 3900.5 & $0.60_{-0.21}^{+0.21}$ & $23.60\pm  0.02$ & $22.77\pm  0.01$ & $22.06\pm  0.01$ &   0.30 & 0.03 \\
1581 & 12:46:15.30 & 30:42:40.8 & 2603.5 & 3888.2 & $0.59_{-0.21}^{+0.21}$ & $24.84\pm  0.03$ & $23.81\pm  0.01$ & $22.88\pm  0.01$ &   0.17 & 0.03 \\
1795 & 12:46:06.38 & 30:43:54.5 & 1222.6 & 1529.8 & $0.59_{-0.21}^{+0.21}$ & $23.50\pm  0.01$ & $23.09\pm  0.01$ & $22.69\pm  0.01$ &   1.98 & 0.00 \\
1821 & 12:46:07.89 & 30:42:05.5 &  581.4 & 3650.3 & $1.38_{-0.31}^{+0.31}$ & $24.37\pm  0.02$ & $24.56\pm  0.02$ & $24.57\pm  0.02$ &   0.38 & 0.03 \\
1877 & 12:46:15.75 & 30:44:59.3 & 3963.2 & 1472.9 & $0.58_{-0.21}^{+0.21}$ & $23.85\pm  0.01$ & $23.35\pm  0.01$ & $22.87\pm  0.01$ &   0.76 & 0.03 \\
2050 & 12:46:13.69 & 30:44:55.0 & 3452.3 & 1307.5 & $0.58_{-0.21}^{+0.21}$ & $23.19\pm  0.02$ & $22.55\pm  0.01$ & $21.98\pm  0.01$ &   2.31 & 0.00 \\
2155 & 12:46:15.47 & 30:45:12.3 & 4017.6 & 1208.7 & $0.69_{-0.22}^{+0.22}$ & $22.89\pm  0.01$ & $22.09\pm  0.01$ & $21.23\pm  0.01$ &   0.25 & 0.03 \\
2211 & 12:46:07.31 & 30:44:21.0 & 1677.7 & 1167.0 & $1.88_{-0.38}^{+0.38}$ & $24.98\pm  0.02$ & $25.25\pm  0.02$ & $25.71\pm  0.04$ &   0.19 & 0.07 \\
2236 & 12:46:03.44 & 30:43:59.4 &  593.0 & 1099.1 & $1.35_{-0.31}^{+0.31}$ & $25.15\pm  0.03$ & $25.28\pm  0.03$ & $25.24\pm  0.03$ &   0.22 & 0.02 \\
2264 & 12:46:13.55 & 30:45:05.1 & 3510.8 & 1110.6 & $0.52_{-0.20}^{+0.20}$ & $23.69\pm  0.02$ & $22.65\pm  0.01$ & $21.91\pm  0.01$ &   0.39 & 0.03 \\
2267 & 12:46:05.14 & 30:44:11.8 & 1095.8 & 1077.6 & $0.55_{-0.20}^{+0.20}$ & $23.68\pm  0.01$ & $23.20\pm  0.01$ & $22.81\pm  0.01$ &   0.38 & 0.03 \\
2413 & 12:46:16.19 & 30:42:30.6 & 2715.6 & 4174.7 & $1.35_{-0.31}^{+0.31}$ & $24.89\pm  0.02$ & $25.11\pm  0.02$ & $25.08\pm  0.02$ &   0.43 & 0.03 \\
2487 & 12:46:05.61 & 30:44:26.4 & 1337.2 &  872.5 & $1.16_{-0.28}^{+0.28}$ & $25.11\pm  0.02$ & $25.21\pm  0.02$ & $24.97\pm  0.02$ &   0.24 & 0.03 \\
2494 & 12:46:14.37 & 30:45:24.5 & 3876.6 &  861.8 & $1.25_{-0.30}^{+0.29}$ & $24.76\pm  0.03$ & $24.89\pm  0.03$ & $24.73\pm  0.03$ &   1.39 & 0.00 \\
2621 & 12:46:15.63 & 30:43:20.3 & 3037.8 & 3221.8 & $1.39_{-0.31}^{+0.31}$ & $24.16\pm  0.02$ & $24.31\pm  0.02$ & $24.35\pm  0.02$ &   0.20 & 0.05 \\
2681 & 12:46:14.85 & 30:45:36.8 & 4098.0 &  699.6 & $0.65_{-0.22}^{+0.22}$ & $25.71\pm  0.04$ & $25.20\pm  0.03$ & $24.59\pm  0.02$ &   0.25 & 0.03 \\
2725 & 12:46:12.46 & 30:45:25.4 & 3444.6 &  622.9 & $0.62_{-0.21}^{+0.21}$ & $21.91\pm  0.01$ & $21.00\pm  0.00$ & $20.20\pm  0.00$ &   0.20 & 0.03 \\
2880 & 12:46:12.79 & 30:45:36.4 & 3622.0 &  465.2 & $0.79_{-0.23}^{+0.23}$ & $25.50\pm  0.05$ & $25.03\pm  0.03$ & $24.16\pm  0.02$ &   1.19 & 0.00 \\
2891 & 12:46:05.87 & 30:44:52.2 & 1630.1 &  443.5 & $0.45_{-0.19}^{+0.19}$ & $22.18\pm  0.01$ & $21.18\pm  0.00$ & $20.55\pm  0.00$ &   3.13 & 0.02 \\
2936 & 12:46:08.67 & 30:45:13.0 & 2461.5 &  399.5 & $0.79_{-0.23}^{+0.23}$ & $23.68\pm  0.02$ & $23.35\pm  0.01$ & $22.64\pm  0.01$ &   0.88 & 0.03 \\
2979 & 12:46:02.79 & 30:44:36.3 &  778.3 &  365.7 & $1.25_{-0.30}^{+0.29}$ & $24.58\pm  0.02$ & $24.66\pm  0.02$ & $24.54\pm  0.02$ &   0.29 & 0.03 \\
3032 & 12:46:01.94 & 30:44:35.0 &  571.6 &  288.9 & $1.15_{-0.28}^{+0.28}$ & $24.18\pm  0.02$ & $24.30\pm  0.02$ & $24.03\pm  0.02$ &   0.86 & 0.01 \\
3043 & 12:46:04.58 & 30:44:52.8 & 1340.2 &  280.3 & $1.41_{-0.32}^{+0.32}$ & $24.65\pm  0.02$ & $24.90\pm  0.02$ & $24.89\pm  0.02$ &   0.55 & 0.03 \\
3052 & 12:46:08.41 & 30:45:19.6 & 2463.5 &  251.4 & $0.73_{-0.23}^{+0.23}$ & $23.30\pm  0.02$ & $22.89\pm  0.01$ & $22.22\pm  0.01$ &   1.94 & 0.00 \\
3084 & 12:46:12.30 & 30:45:47.2 & 3605.9 &  214.9 & $0.54_{-0.20}^{+0.20}$ & $21.72\pm  0.01$ & $21.22\pm  0.01$ & $20.83\pm  0.01$ &   0.44 & 0.03 

\enddata
\tablenotetext{a}{SExtractor ID}
\tablenotetext{b}{Right Ascension (J2000)}
\tablenotetext{c}{Declination (J2000)}
\tablenotetext{d}{Bayesian photometric redshift}
\tablenotetext{e}{AB Isophotal magnitude in the F475W filter}
\tablenotetext{f}{AB Isophotal magnitude in the F606W filter}
\tablenotetext{g}{AB Isophotal magnitude in the F814W filter}
\tablenotetext{h}{Full width at half maximum as measured by $SExtractor$ in arcsec}
\tablenotetext{i}{SExtractor star/galaxy classification}
\tablecomments{Catalog with magnitudes and photometric redshifts in the NGC 4676 field. Only
galaxies outside the masked area with $I_{F814W} < 26$, and very high values of the Bayesian odds ($>0.99$) are
included. This is the subsample of galaxies for which the photometric redshifts are most reliable.
The full catalog published electronically contains more information about these objects and about
the rest of the detections in the field.}
\end{deluxetable}

\clearpage

\appendix

\centerline{\bf{Appendix}}

\section{$BUCS$ simulations}

A robust, model-independent way of generating realistic galaxy fields is to 
take deep observations already available and then rearrange the objects to generate 
another field.  Using this approach, we simulate deep ACS images with the $BUCS$ 
(Bouwens Universe Construction Set) software.  In the first step, we 
determine the number of times each object appears in a
given image by drawing from a Poisson distribution with mean \( \sigma
_{obj}A_{sim} \), where \( \sigma _{obj} \) is the surface density of
the object and \( A_{sim} \) is the field area being simulated.  In the
second step, we distribute the objects across the field assuming a uniform
random distribution, i.e. no spatial clustering. In the third step,
we simulate ACS images in any number of filters using the Monte-Carlo
catalogs generated in the first two steps. We place objects in these
images in one of two ways: using their best-fit analytic profiles or
resampling the original objects onto the image. To do this properly,
BUCS (1) k-corrects each object template using best-fit pixel-by-pixel
and object SED and (2) corrects its PSF to match the PSF for the ACS
filter being simulated. Finally, we add the expected amount of noise
to the image. The formalism used to perform these final two steps is
described more extensively in Bouwens, Broadhurst, and Illingworth
(2003) and Bouwens (2003, in preparation).

Since these simulations are just a rearrangement of objects from an
input field, they should be an extremely accurate representation of
the observations, not only in number, angular size, ellipticity
distributions, and color distributions, but also in morphology and
pixel-by-pixel color variations. 

Because BUCS is an extremely complex set of software, a more
user-friendly interface, $BUCS\_LITE$, has been written. The software can be 
downloaded from: \break 
{\noindent http://www.ucolick.org/$\sim$bouwens/bucs/index.html.}  

The relevant parameters used for the $BUCS\_LITE$ simulations described
in this paper are given in Table A8.

\begin{deluxetable*}{lc}
\tabletypesize{\scriptsize}
\tablecaption{Parameters for $BUCS\_LITE$ v1.0b (used with $BUCS$ v1.0)} 
\tablewidth{0pt} 
\tablehead{
\colhead{Parameter} & 
\colhead{Value} } 
\startdata 
$MAG\_MIN$ & 15.000 	\cr
$MAG\_MAX$ & 30.000	\cr
$FILTER\_FILE$ 	& ACS\$wfc/filter.struct	\cr
$PROFILE\_TYPE$ 	& ALL	\cr
$TEMPLATES$ 	& HDF\_Analytic (Tadpole\_Real for the deblending
simulations)
\enddata
\end{deluxetable*}

\section{Relevant $SExtractor$ Configuration Parameters}

\subsection{Detection and deblending}
The SExtractor parameters used for detection and deblending of the galaxies are given in Table B9.

\begin{deluxetable*}{lc}
\tabletypesize{\scriptsize}
\tablecaption{Parameters used for the Detection and Deblending with SExtractor v2.2.2}
\tablewidth{0pt} 
\tablehead{
\colhead{Parameter} & 
\colhead{Value} } 
\startdata
$BACK\_FILTERSIZE$ 	&  5		\cr
$BACK\_SIZE$ 		&  128	\cr
$FILTER$ 			&  Y		\cr
$FILTER\_NAME$ 		&  gauss\_2.0\_5x5.conv  \cr
$WEIGHT\_TYPE$ 		&  MAP\_WEIGHT	\cr
$WEIGHT\_THRESH$ 		&  0, 1.0e30	\cr
$INTERP\_TYPE$ 		&  NONE	\cr
$DETECT\_MINAREA$ 	&  5		\cr
$DETECT\_THRESH$ 		&  1.5	\cr
$DEBLEND\_NTHRESH$ 	&  16		\cr
$DEBLEND\_MINCONT$ 	&  0.025	\cr
$CLEAN$ 			&  Y		\cr
$CLEAN\_PARAM$ 		&  1.2	
\enddata
\end{deluxetable*}

\subsection{Photometry and analysis}

The parameters used for the SExtractor photometry and analysis are listed in Table B10.

\begin{deluxetable*}{lc}
\tabletypesize{\scriptsize}
\tablecaption{Parameters used for Photometry and Analysis with SExtractor v2.2.2}
\tablewidth{0pt} 
\tablehead{
\colhead{Parameter} & 
\colhead{Value} } 
\startdata
$ANALYSIS\_THRESH$ 	&  1.5	\cr
$BACKPHOTO\_TYPE$ 	&   LOCAL	\cr
$BACKPHOTO\_THICK$ 	&   26	\cr
$MASK\_TYPE$ 		&   CORRECT	\cr
$PHOT\_APERTURES$ &   2,3,4,6,8,10,14,20,28,40,60,80,100,160 (diameters) \cr
$PHOT\_AUTOPARAMS$ 	&   2.5, 3.3	\cr
$PIXEL\_SCALE$ 	&   0.05	\cr
$GAIN$ 		&   1.0	\cr
$PHOT\_FLUXFRAC$ 	&   0.5, 0.9	\cr
$STARNNW\_NAME$ 	&   default.nnw	\cr
$SEEING\_FWHM$ 	&   0.105	
\enddata
\end{deluxetable*}

\section{BPZ parameters}

The photometric redshifts in this paper were calculated with the BPZ parameters listed in Table C11 (only those different from the defaults are given in the table).

\begin{deluxetable*}{lc}
\tabletypesize{\scriptsize}
\tablecaption{Non-default parameters used by BPZ v1.98b}
\tablewidth{0pt} 
\tablehead{
\colhead{Parameter} & 
\colhead{Value} } 
\startdata
$SPECTRA$ 	&  CWWSB\_Benitez2003.list	\cr
$INTERP$ 	&  2		\cr
$DZ$ 		&  0.002	
\enddata
\end{deluxetable*}

\end{document}